\documentclass[modern, letterpaper]{aastex61}

\usepackage{amsmath}
\usepackage{amsfonts}
\usepackage{amssymb}
\usepackage{booktabs}

\usepackage{graphicx}
\usepackage{color}

\usepackage{hyperref}
\definecolor{linkcolor}{rgb}{0.02,0.35,0.55}
\definecolor{citecolor}{rgb}{0.4,0.4,0.4}
\hypersetup{colorlinks=true,linkcolor=linkcolor,citecolor=citecolor,
            filecolor=linkcolor,urlcolor=linkcolor}
\hypersetup{pageanchor=false}

\newcommand{\documentname}{\textsl{Article}}
\newcommand{\sectionname}{Section}

\newcommand{\eqname}{Equation}

\newcommand{\package}[1]{\texttt{#1}}
\newcommand{\acronym}[1]{{\small{#1}}}

\newcommand{\python}{\package{Python}}

\newcommand{\project}[1]{\textsl{#1}}

\newcommand{\transpose}[1]{{#1}^{\!\mathsf{T}}}

\renewcommand{\vec}[1]{\bs{#1}}
\newcommand{\mat}[1]{\mathbf{#1}}

\newcommand{\kms}{\ensuremath{\mathrm{km}~\mathrm{s}^{-1}}}
\newcommand{\au}{\ensuremath{\mathrm{au}}}
\newcommand{\pc}{\ensuremath{\mathrm{pc}}}

\newcommand{\bs}[1]{\boldsymbol{#1}}

\usepackage{microtype}
\usepackage{multirow}

\newcommand{\gaia}{\project{Gaia}}
\newcommand{\DR}[1]{\acronym{DR}#1}
\newcommand{\tgas}{\acronym{TGAS}}
\newcommand{\tmass}{\acronym{2MASS}}
\newcommand{\lbfgsb}{\texttt{L-BFGS-B}}
\newcommand{\Ha}{\ensuremath{{\rm H}\alpha}}

\newcommand{\logg}{\ensuremath{\ensuremath{\log g}}}
\newcommand{\teff}{\ensuremath{\ensuremath{T_{\rm eff}}}}
\newcommand{\llrold}{\ensuremath{\mathcal{R}_\mu}}
\newcommand{\llrnew}{\ensuremath{\mathcal{R}_{\rm RV}}}

\newcommand{\npairsobs}{311}
\newcommand{\ncomoving}{127}
\newcommand{\nraveoverlap}{154}
\newcommand{\nravecomoving}{59}
\newcommand{\ncomovingtotal}{186}

\definecolor{mediumpersianblue}{rgb}{0.0, 0.4, 0.65}

\begin{document}\sloppy\sloppypar\raggedbottom\frenchspacing 

\title{Spectroscopic confirmation of very-wide stellar binaries and
       large-separation comoving pairs from \textsl{Gaia DR1}}

\author[0000-0003-0872-7098]{Adrian~M.~Price-Whelan}
\affiliation{Department of Astrophysical Sciences,
             Princeton University, Princeton, NJ 08544, USA}

\author[0000-0001-7790-5308]{Semyeong~Oh}
\affiliation{Department of Astrophysical Sciences,
             Princeton University, Princeton, NJ 08544, USA}

\author[0000-0002-5151-0006]{David~N.~Spergel}
\affiliation{Flatiron Institute,
             Simons Foundation,
             162 Fifth Avenue,
             New York, NY 10010, USA}
\affiliation{Department of Astrophysical Sciences,
             Princeton University, Princeton, NJ 08544, USA}



\correspondingauthor{Adrian M. Price-Whelan}
\email{adrn@astro.princeton.edu}

\begin{abstract}
Members of widely-separated stellar binaries ($\gtrsim 0.1$~pc) or loosely bound
stellar associations and clusters are born with small spreads in kinematics and
chemistry, but disrupt and dissociate on timescales comparable to their orbital
time within the Milky Way.
The kinematic properties of comoving star pairs---bound wide binaries or unbound
but coeval stars---are therefore sensitive to both the smooth tidal field and
gravitational perturbation spectrum of the Milky Way.
In previous work, we identified $>4,000$ candidate comoving pairs using only
astrometric data from the Tycho-Gaia Astrometric Solution (TGAS) catalog and
find a large number of pairs with separations $\gtrsim 1~\pc$.
Without radial velocity measurements for the vast majority of these pairs, the
false-positive rate at these separations is significant ($\sim 50\%$).
Here we present results from our own low-resolution radial-velocity survey of a
random sample of the candidate pairs to identify and validate true comoving star
pairs.
Of the \npairsobs\ observed comoving pairs, we confirm \ncomoving\ comoving
pairs with separations as large as $\approx 10~\pc$, our original search limit.
At separations of $10^{-3}~\pc$ to $\approx 0.5~\pc$,
the number of confirmed pairs per separation decreases with increasing
separation (number per log separation is uniform).
From $\approx 0.5~\pc$ to $10~\pc$, the number per separation is approximately
uniform (number per log separation increases).
We confirm the discovery of a population of comoving star pairs at very large
separations, suggesting that disrupted wide binaries and stellar associations
remain approximately spatially coherent.
\end{abstract}



\section{Introduction}\label{sec:introduction}

Widely-separated, coeval star pairs or multiples --- i.e. wide binaries,
multiplets, or former binaries or multiplets --- are important tracers of many
processes that formed and continue to shape the Milky Way.
They have been used to place limits on the mass distribution and population of
massive perturbers in the Galaxy (e.g.,
\citealt{Yoo:2004,Monroy-Rodriguez:2014}), to study the physics of star
formation (e.g., \citealt{Parker:2009,Reipurth:2012}), and to calibrate stellar
models and inferences across the main sequence (e.g., \citealt{Brewer:2016}).
These examples represent a small subset of a much broader range of applications
(see, e.g., \citealt{Chaname:2007}).

At smaller separations ($\lesssim 0.1~\pc$), comoving pairs of stars are likely
bound wide binaries.
It has long been recognized that the observed period or separation distribution
of these low-binding-energy binaries encapsulates information about the
Galactic mass distribution (\citealt{Opik:1924,Oort:1950,Bahcall:1985}).
In particular, early simulations of populations of wide binaries showed that
the expected separation distribution, $p(s)$, for $0.01 < s < 1~\pc$ tends to
follow a power law, $p(s) \propto s^{-\alpha}$ for $\alpha \approx 1$--$2$,
with a possible steep break that depends on the details of the population of
massive perturbers in the Galaxy, i.e. giant molecular clouds (GMCs), black
holes, etc. (\citealt{Weinberg:1987}).

At larger separations ($\gtrsim 0.1~\pc$), comoving pairs are likely a mix of
tenuously bound stars such as recently disrupted binaries, moving groups or resonant
features, and the final products of dissolving stellar associations or
clusters.
However, even at separations of $\sim 0.1~\pc$, we typically lack the velocity
or distance precision to determine the binding energy of a system and therefore
can't determine its true dynamical state; because of this ambiguity, we refer to
both bound and unbound widely-separated ($\gtrsim 10^{-3}~\pc$) systems
collectively as ``comoving'' systems.

Primarily thanks to large astrometric surveys and catalogs
(\citealt{ESA:1997,Lepine:2005,Gaia-Collaboration:2016}) and radial-velocity
surveys (\citealt{Steinmetz:2006}), several hundred confirmed comoving star
pairs (\citealt{Shaya:2011}) and over 4,000 candidate systems
(\citealt{Gould:2004,Lepine:2007,Tokovinin:2012,Allen:2014,Oh:2017,
Oelkers:2017, Andrews:2017}) have been discovered over the last several decades
(see also \citealt{Chaname:2007} and references therein).
These pairs are identified by searching for approximately cospatial stars that
appear to be comoving either in sky-projected velocities --- we refer to these
as ``comoving pair candidates'' --- or using full-space velocity information ---
we refer to these as ``confirmed comoving pairs.''
Radial velocity follow-up is still needed to confirm the majority of the known
candidates, though \gaia\ \DR{2} will provide radial velocities for the
brightest stars ($G < 12$; \citealt{Gaia-Collaboration:2016}).

With large samples of candidate pairs, several groups have measured the
separation distribution of wide binaries at angular separations $\Delta \theta
\lesssim 1^\circ$ where the false-positive rate is expected to be lower
and the samples are likely predominantly wide binaries
(\citealt{Chaname:2004,Lepine:2007,Sesar:2008}).
In most cases, the angular or projected separation distribution is fit using a
decreasing power-law, and the power-law index, $\alpha$, is found to be $\alpha
\approx 1.5$--$2$ out to angular separations of $\approx 0.25^\circ$ ($\approx
0.65~\pc$ at a distance of $150~\pc$), apparently consistent with simulations
(e.g., \citealt{Weinberg:1987}).
Tentative reports of a steepening or cutoff of the separation distribution at
larger separations have been claimed (e.g., \citealt{Yoo:2004,Quinn:2009}), but
these claims are limited by small numbers of confirmed pairs at large
separations.

Most measurements of the separation distribution have neglected the possible
existence of larger separation ($\gtrsim 1~\pc$) unbound or ionized pairs.
Early theoretical work either explicitly ignored these stars (e.g.,
\citealt{Weinberg:1987}), or suggested that unbound stars would drift and no
longer remain spatially coherent within one Galactic orbit (e.g., see Appendix
in \citealt{Yoo:2004}).
More recent simulations of the disruption and evolution of wide binary
populations include an approximation of the Galactic tidal field and instead
predict that ionized binaries could appear as a second peak or component in the
separation distribution of comoving stars at much larger separations
(10--$1000~\pc$; \citealt{Jiang:2010}).
Though these simulations don't include perturbers like GMCs or black holes, they
at least indicate that disrupted wide binaries may remain coherent for a longer
period of time than previously thought.
Thus, a monotonically decreasing separation distribution may be too simple of a
model for the separation distribution of comoving stars in our Galaxy.
In addition, the distribution of unbound but comoving stars should be very
sensitive to diffusive processes that determine the rate at which the pairs
separate.

Indeed, observational work that followed appears to find tentative evidence for
the existence of a second peak in the separation distribution of
radial-velocity-confirmed comoving stars (\citealt{Shaya:2011}; hereafter SO11).
A key aspect of this work was to move from searching for common proper motion
pairs to performing a statistical inference of the difference in full-space
velocity vector for pairs of stars.
At separations of $1$--$10~\pc$, star pairs in the local solar neighborhood
subtend large angles on the sky: even if the two component stars were moving
with the same 3D velocity, their proper motions and radial velocities can be
different.
By compiling radial velocities from the literature for \project{Hipparcos}
stars, SO11 applied this method to search for all comoving star pairs with
separations larger than $0.01~\pc$ within $100~\pc$ of the sun.
They find $\sim 200$ high probability comoving pairs with a separation
distribution that falls of until $\sim 1~\pc$, where the number of pairs then
begins rising in bins of log-separation.

To verify the existence of a population of very wide-separation comoving pairs,
we first need to identify a large sample of confirmed comoving stars.
This will already be possible for the brightest stars with \gaia\ \DR{2}, but
obtaining even larger samples will require significant efforts to obtain radial
velocities for candidate comoving pairs.
In this work, we present a catalog of confirmed comoving star pairs discovered
through low-resolution spectroscopic follow-up of candidate comoving stars
selected from inferences using astrometric measurements from the \tgas\ catalog
alone (\citealt{Oh:2017}).

We describe the data and observations in \sectionname~\ref{sec:data}.
Details of the spectroscopic reduction and velocity measurements are described
in the Appendix.
We present our main results in \sectionname~\ref{sec:results} and
\sectionname~\ref{sec:discussion}.

\section{Data}\label{sec:data}

\begin{figure}[htb!]
  \begin{center}
    \includegraphics[width=0.8\linewidth]{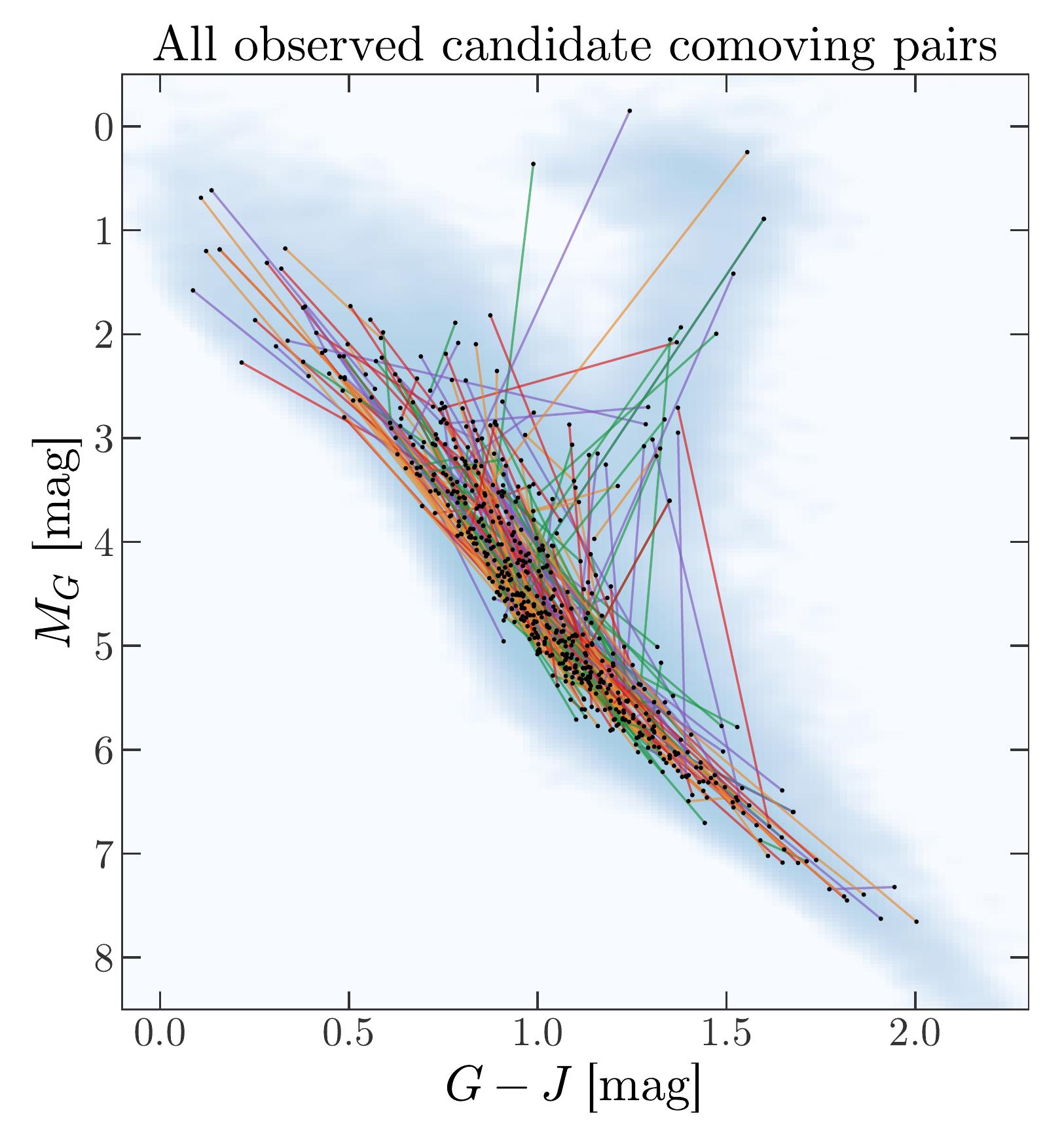}
  \end{center}
  \caption{%
    Color-magnitude diagram for all \tgas\ sources within $200~\pc$ (background
    blue histogram) and all comoving star pair members observed in this
    follow-up effort (black points); colored lines connect candidate comoving
    star pairs.
    \tmass\ (\citealt{Skrutskie:2006}) $J$-band magnitudes are obtained
    from the Gaia science archive.
    Apparent $G$-band magnitudes (\citealt{Carrasco:2016}) are converted to
    absolute magnitude, $M_G$, using a Lutz-Kelker-corrected
    (\citealt{Lutz:1973}) distance estimate (see also \eqname~1 in
    \citealt{Oh:2017}).
    \label{fig:sample-cmd}}
\end{figure}

\subsection{Sample selection}\label{sec:sample}

In previous work, we have identified a sample of high-confidence, candidate
comoving star pairs (\citealt{Oh:2017}) using only astrometric information from
the \tgas\ catalog (\citealt{Michalik:2015,Gaia-Collaboration:2016a}).
These comoving pairs were selected using a cut on a marginalized likelihood
ratio computed for all nearly co-spatial pairs of stars.
We performed a model selection between two hypotheses: that the velocities of
the two stars in a pair are identical (case 1) or that they are indepndent (case
2).
We marginalized over the (unknown) true velocity and distance of each star to
obtain the ratio of fully marginalized likelihoods,
$\mathcal{L}_1/\mathcal{L}_2$.
The likelihoods appropriately take into account the reported uncertainties and
covariances of the astrometric measurements in the \tgas\ catalog
as well as projection effect in comparing spherical
velocities for star pairs separated by large angles on the sky.

After a parallax signal-to-noise ratio (SNR) cut, $\varpi/\sigma_\varpi > 8$,
we compute the marginalized likelihood ratio for all stars within $10~\pc$ and
with a difference in tangential velocity $\Delta v_\perp < 10~\kms$ for each
star in the \tgas\ catalog.
After applying a conservative cut on the likelihood ratio motivated by computing
the same likelihood ratio for random pairings of stars, $\ln
\mathcal{L}_1/\mathcal{L}_2 > 6$, we found 13,058 comoving star pairs with
physical separations between $10^3~\au$ and $10~\pc$.
Of these pairs, there are 10,606 unique sources, implying that a significant
fraction of the identified pairs are members of larger groups.
However, we have also found a large number ($\approx$4,000) of
mutually-exclusive linked pairs that increase in number with increasing
log-separation from $\sim 1$~\pc.
For a full description of the selection method, see \citealt{Oh:2017}.

To select candidate pairs for follow-up spectroscopic observations, we only
consider mutually exclusive star pairs within $200~\pc$ in Heliocentric
distance.
With the SNR cut used above and given the limitations of the \tgas\ catalog
itself, \tgas\ is $\approx$90\% complete to $200~\pc$ (\citealt{Bovy:2017}).
From this source target list, we randomly observed targets with airmass $\sec z
< 1.5$ during our observing run.
\figurename~\ref{fig:sample-cmd} shows a color-magnitude diagram of all targets
observed in this observing run (markers) plotted over the density of all \tgas\
stars within $200~\pc$.
Candidate comoving pairs are connected by (colored) lines.

\subsection{Observations and data reduction}\label{sec:reduction}

Spectra were obtained using the \project{Modspec} spectrograph mounted on the
$2.4~{\rm m}$ Hiltner telescope at MDM
observatory\footnote{\url{http://mdm.kpno.noao.edu}} on Kitt Peak (Arizona).
\project{Modspec} was set up in long-slit mode with a 1\arcsec\ slit, a
$600~{\rm line}~{\rm mm}^{-1}$ grating, and a $2048\times2048~{\rm pixel}$ SITe
CCD detector (``Echelle'').
Only the central 300 pixel columns were read out from the CCD, and pixel rows
$>1600$ (at the blue end) were later trimmed because of the poor quantum
efficiency of the CCD at wavelengths $\lesssim 3600~{\rm \AA}$.
The resolution and wavelength range were $\approx2~{\rm \AA}~{\rm pixel}^{-1}$
from $\approx 3600$--$7200~{\rm \AA}$.
At this resolution, to obtain high signal-to-noise ($>100$ per pixel) spectra,
most exposures were between $30$--$120~{\rm s}$.
To maximize the number of observed pairs, we chose not to take
comparison lamp spectra at each pointing as it adds an overhead of $\approx
90~{\rm s}$ per observation.
We are primarily interested in relative velocities between pair members and
therefore don't need precise absolute calibrations of their radial velocities.
To determine rough nightly wavelength solutions, calibration spectra were
obtained using Hg-Ne and Ne comparison lamps at the beginning and end of each
night.
Additionally, atmospheric emission lines were used to correct the wavelength
solution on a per-object basis.

We observed a total of 765 unique sources over 5 nights from (UT) 2017-03-11 to
2017-03-15, corresponding to 323 unique pairs and $\approx 120$
calibration targets.
All calibration targets have previously measured radial velocities from
high-resolution spectroscopic surveys (\citealt{Bensby:2014,Pinsonneault:2014})
in order to validate our radial velocity measurements.
These stars also have measured chemical abundances, and will be used in future
work as a training set for \project{The Cannon} (\citealt{Ness:2015}) to infer
metallicities and stellar parameters for our observed sources.
The spectra were reduced and calibrated using a custom, publicly-available
pipeline written in \python; see Appendices~\ref{sec:1d-ext}--\ref{sec:rv-meas}
for a full description of the data reduction procedure.

\subsection{Validation of absolute radial velocities}\label{sec:rv-validation}

\begin{figure}[!ht]
  \begin{center}
    \includegraphics[width=0.85\linewidth]{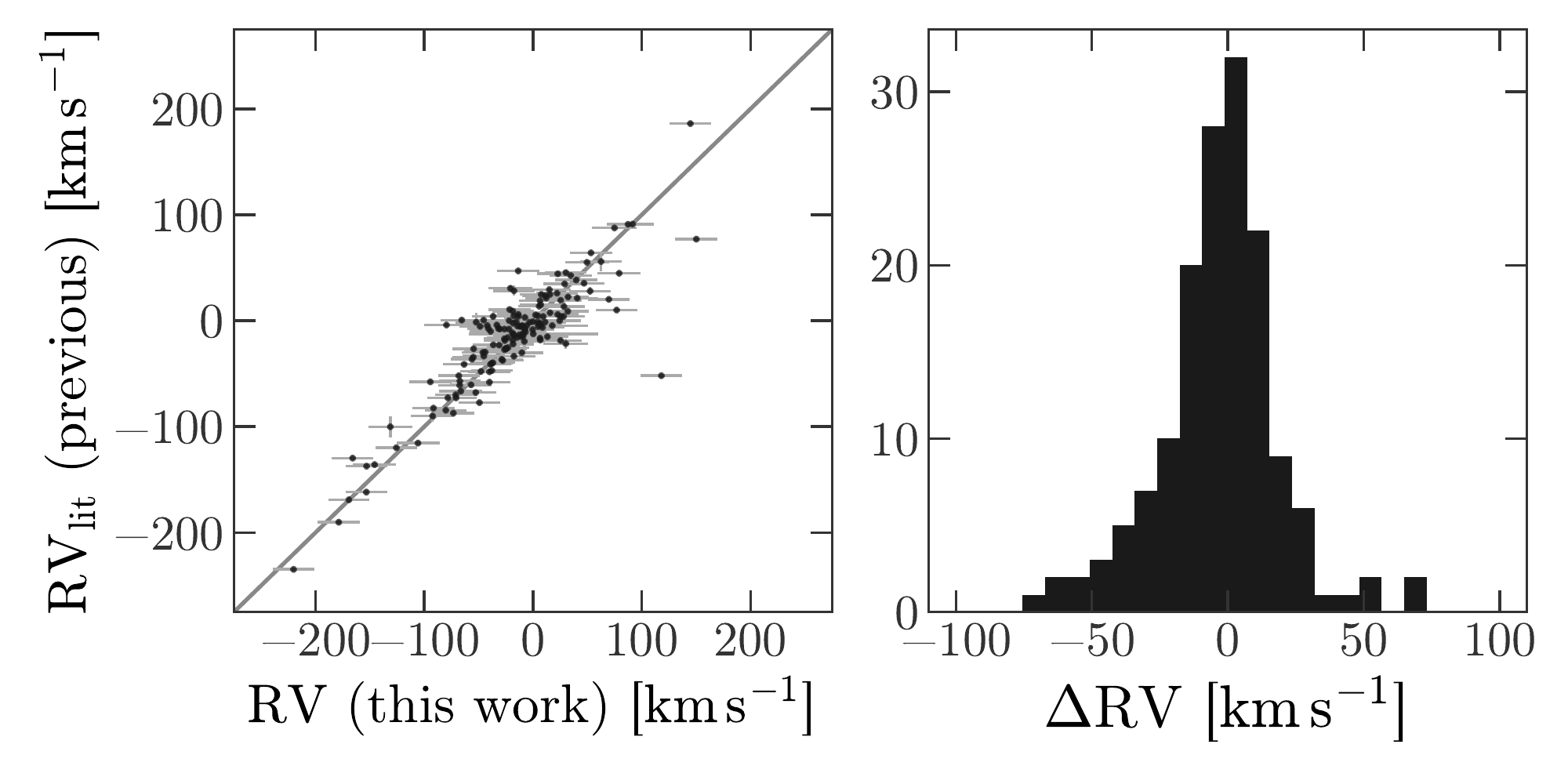}
  \end{center}
  \caption{%
    Comparison of radial velocities measured from our observations with radial
    velocities obtained from the literature search described in
    \sectionname~\ref{sec:rv-validation}.
    Left panel shows our values on the $x$-axis, ${\rm RV}$, against the
    previous literature values on the $y$-axis, ${\rm RV}_{\rm lit}$, for 155
    observed stars (mainly calibration targets) with previously measured radial
    velocities.
    Right panel shows a histogram of the differences, $\Delta{\rm RV} = {\rm RV}
    - {\rm RV}_{\rm lit}$.
    The estimated standard-deviation of the difference distribution implies a
    global systematic uncertainty of $\approx 20~\kms$.
    \label{fig:compare-previous}}
\end{figure}

Several of our targets and all of our calibration targets have previously
measured radial velocities (RVs).
We can therefore estimate the absolute velocity uncertainties by comparing our
derived radial velocities with high-quality prior measurements.
For calibration sources, we use radial velocity values reported in the source
catalogs (\citealt{Bensby:2014,Pinsonneault:2014}).
For all other sources, following previous work (\citealt{Shaya:2011}), we search
for previous RV measurements (in decreasing order of priority) with:
\begin{enumerate}
  \item The Geneva-Copenhagen Survey of the Solar Neighborhood
  (\citealt{Nordstrom:2004}).
  \item The Catalogue of Radial Velocities of Galactic Stars with Astrometric
  Data, the Second Version (\citealt{Kharchenko:2007}).
  \item The General Catalogue of Mean Radial Velocities
  (\citealt{Barbier-Brossat:2000}).
  \item The SIMBAD database (\citealt{Wenger:2000}). We use \package{astroquery}
  (\citealt{Ginsburg:2016}) to retrieve records from the database, which do not
  contain uncertainties but do contain a quality flag. We only retain RVs that
  have a quality flag set to `A' and adopt a $5~\kms$ uncertainty for all such
  values.
\end{enumerate}
We find that 155 of the observed stars (mainly calibration targets) have
reliable prior measurements from the literature search described above.
\figurename~\ref{fig:compare-previous} shows the distribution of velocity
differences between our measurements and literature values.
From this distribution (right panel), we compute the median absolute deviation
(MAD) and estimate the standard deviation to be $\approx 20~\kms$, which we
adopt as our global systematic error for all absolute radial velocities.
We note, however, that for computing \emph{relative} velocities for pairs of
nearby stars, it is the RV precision (typically $\approx 5$--$10~\kms$) that
matters.

\section{Results} \label{sec:results}

\subsection{Comoving pair catalog}\label{sec:catalog}

Of the 323 observed star pairs, 12 have at least one member with an unreliable
(flagged) radial velocity measurement from the spectroscopic reduction pipeline.
For the remaining 311 pairs, we use the measured radial velocities to re-compute
the marginal likelihood ratio for the two hypotheses considered in previous work
(see Appendix~B, \citealt{Oh:2017}).
We compute two fully marginalized likelihoods (FMLs) for the data: (1) assuming
that the two stars are comoving with identical 3D velocity vectors (within some
allowed small dispersion), and (2) assuming that the two stars have independent
3D velocity vectors drawn from a prior velocity distribution.
We set the small allowed dispersion ($s$, in previous work) to $1~\kms$, and use
a mixture of three Gaussian velocity components for the prior, meant to
represent the thin disk, thick disk, and halo velocity distributions relative to
the sun (see \citealt{Oh:2017}).

To compute the new likelihood ratio, we modify the expressions from
\citet{Oh:2017} to include the radial velocity measurements.
In the same notation, this requires the following changes:
\begin{itemize}
  \item The projection matrix, $\mat{M}$, now has to include the radial unit
    vector
    \begin{equation}
      \mat{M} =
        \left(
          \begin{array}{c c c}
            -\sin\alpha & \cos\alpha & 0 \\
            -\sin\delta \, \cos\alpha & -\sin\delta \, \sin\alpha & \cos\delta\\
            \cos\delta \, \cos\alpha & \cos\delta \, \sin\alpha & \sin\delta
          \end{array}
        \right)
    \end{equation}
    where $(\alpha, \delta)$ are the ICRS coordinates (right ascension and
    declination) of a given star.
  \item The data-space velocity vector, $\vec{y}$, for one star now becomes
    \begin{equation}
      \vec{y} =
        \transpose{\left(
          \begin{array}{c c c}
            r_i\,\mu_{\alpha}^* &
            r_i\,\mu_{\delta} &
            v_{r}
          \end{array}
        \right)}
    \end{equation}
    where $\mu_\alpha^*$ is the proper motion in right ascension (including the
    $\cos{\delta}$ term), $\mu_\delta$ is the proper motion in declination, and
    $v_r$ is the radial velocity.
  \item The covariance matrix becomes the block-diagonal matrix
    \begin{equation}
      \mat{\Sigma} = \left(
        \begin{array}{c c}
          r^2 \, \mat{C} & 0 \\
          0 & \sigma_{v_r}^2
        \end{array}
      \right)
    \end{equation}
    where $r$ is the distance, $\mat{C}$ is the proper motion covariance matrix
    (from the \tgas\ catalog), and $\sigma_{v_r}$ is the radial velocity
    uncertainty.
\end{itemize}
To compute the FML for hypothesis 1 for a pair of stars $(i,j)$, the data-space
vector must become the 6-vector
\begin{equation}
  \vec{y} = \transpose{\left(
    \begin{array}{c c}
      \vec{y}_i &
      \vec{y}_j
    \end{array}
    \right)}
\end{equation}
and the covariance matrix becomes the block-diagonal matrix
\begin{equation}
  \mat{\Sigma} = \left(
    \begin{array}{c c}
      \mat{\Sigma}_i & 0 \\
      0 & \mat{\Sigma}_j
    \end{array}
  \right) \quad .
\end{equation}

\begin{figure}[htbp]
  \begin{center}
    \includegraphics[width=0.5\linewidth]{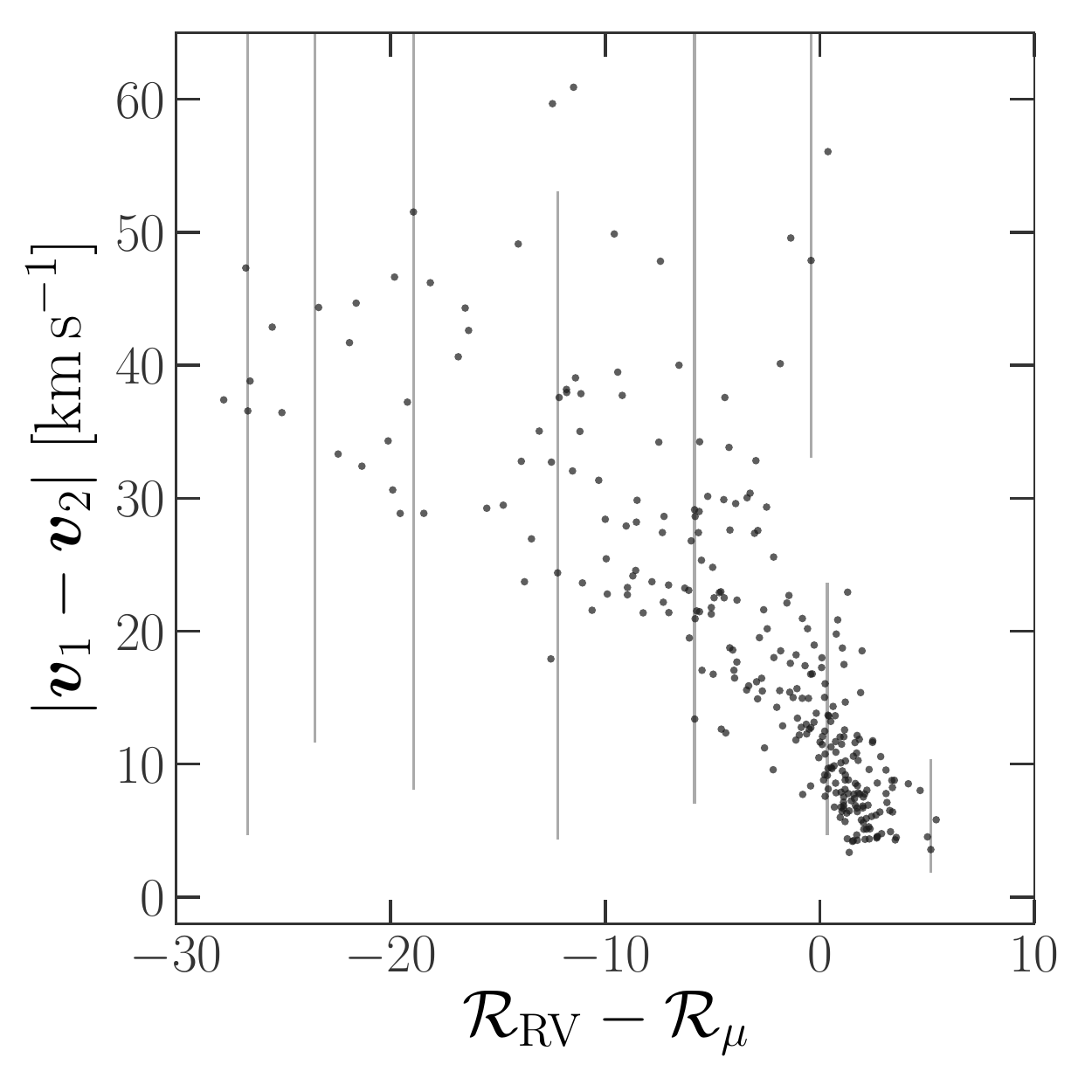}
  \end{center}
  \caption{%
    Point estimate of the 3D velocity difference between two stars in a comoving
    pair, $|\bs{v}_1 - \bs{v}_2|$, plotted against the change in
    log-likelihood-ratio computed when including radial velocity data.
    Error bars are shown for a few randomly-chosen, representative pairs and
    represent the 15th and 85th percentiles of velocity differences computed
    from samples from the error distributions of parallax, proper motion
    components, and radial velocity.
    The likelihood ratio increases (i.e. the belief that the stars are comoving
    becomes stronger) when the 3D velocity difference is small.
    \label{fig:llr}}
\end{figure}

If the new likelihood ratio is larger than that computed from astrometry alone,
the added radial velocity information supports the belief that the two stars are
truly comoving in 3-space.
Conversely, if the new likelihood ratio is smaller, the radial velocity data
indicate that the pair is likely a false-positive and are not truly comoving.
In practice, we compute the logarithm of the likelihood ratios,
$\mathcal{R} = \ln \mathcal{L}_1/\mathcal{L}_2$; we refer to the
proper-motion-only log-FML-ratio for a given pair of stars as
\llrold\ (as computed in \citealt{Oh:2017}) and the log-FML-ratio including the
radial velocity data as \llrnew.
We consider a pair as being genuinely comoving if $\llrnew > \llrold$:
\ncomoving\ of the \npairsobs\ comoving pairs satisfy this constraint.
\figurename~\ref{fig:llr} shows the magnitude of the 3D velocity difference
plotted against the change in log-likelihood-ratio, $\llrnew - \llrold$, for all
observed pairs.
In detail, for visualization, we sample from the error distributions over proper
motion, parallax, and radial velocity for each star individually and compute the
magnitude of the 3D velocity differences, $|\bs{v}_1-\bs{v}_2|$, for $2^{16}$
samples.
We plot the median of the resulting distribution of velocity-differences as
dark markers (circles).
The error bars show the 15th and 85th percentiles of the distributions for a few
randomly chosen pairs.
Note that pairs with large 3D velocity differences have large (negative) changes
to their log-likelihood-ratio values.
Some pairs with large velocity differences also have small (near zero) changes
to the log-likelihood ratio: these are pairs where each member has a velocity
that is an outlier with respect to the prior velocity distribution.

The inferred 3D velocity differences and values for \llrnew\ and \llrold\ for
all observed candidate comoving pairs are provided with the associated table
(\tablename~\ref{tbl:data-pairs}; see Appendix~\ref{sec:table}).

\begin{figure}[htb]
  \begin{center}
    \includegraphics[width=0.9\linewidth]{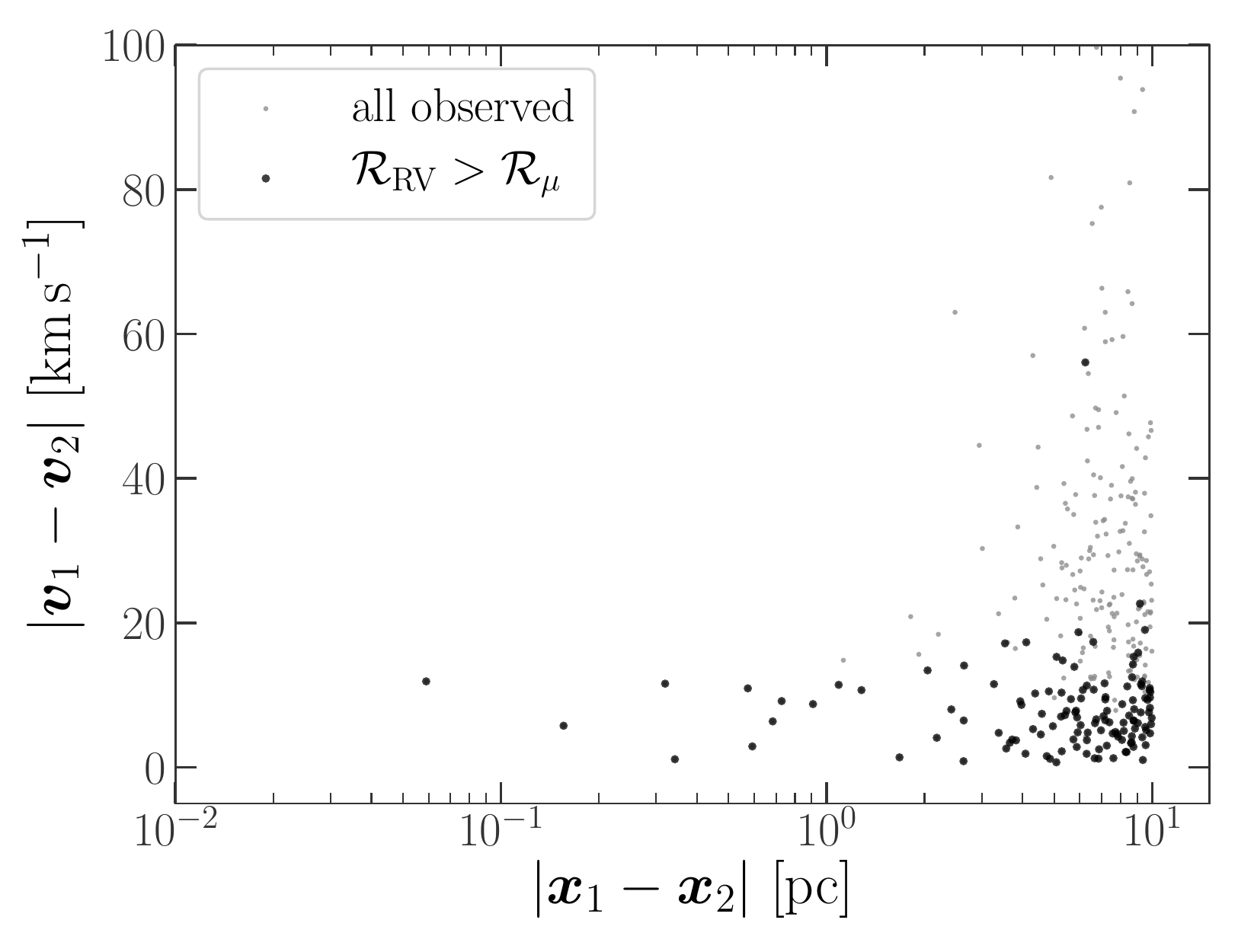}
  \end{center}
  \caption{%
    Point estimate of the 3D velocity difference between two stars in a comoving
    pair, $|\bs{v}_1 - \bs{v}_2|$, plotted against a point estimate of the 3D
    spatial separation of the two stars, $|\bs{x}_1 - \bs{x}_2|$.
    The distances used in the point estimates are computed using a
    Lutz-Kelker-corrected parallax value (\citealt{Lutz:1973}).
    \label{fig:dx-dv}}
\end{figure}

\subsection{Confirmed comoving pairs}\label{sec:genuine}

\figurename~\ref{fig:dx-dv} shows the magnitude of the 3D velocity difference
plotted against the 3D positional separation of the two stars in all observed
pairs from this work.
We again plot the median of the velocity difference distribution, and here use
the Lutz-Kelker-corrected distance estimate (\citealt{Lutz:1973}) computed from
the stars' \tgas\ parallax measurements to compute the 3D positions of the two
stars.
Small, gray markers are pairs for which $\llrnew < \llrold$, i.e. where the
radial velocity measurements weaken or change our belief that a given pair is
comoving.
Larger, black markers are confirmed comoving pairs ($\llrnew > \llrold$); note
that many large-separation pairs have been ruled-out as comoving, but
most small-separation pairs maintain their status as comoving.

\begin{figure}[htb]
  \begin{center}
    \includegraphics[width=\linewidth]{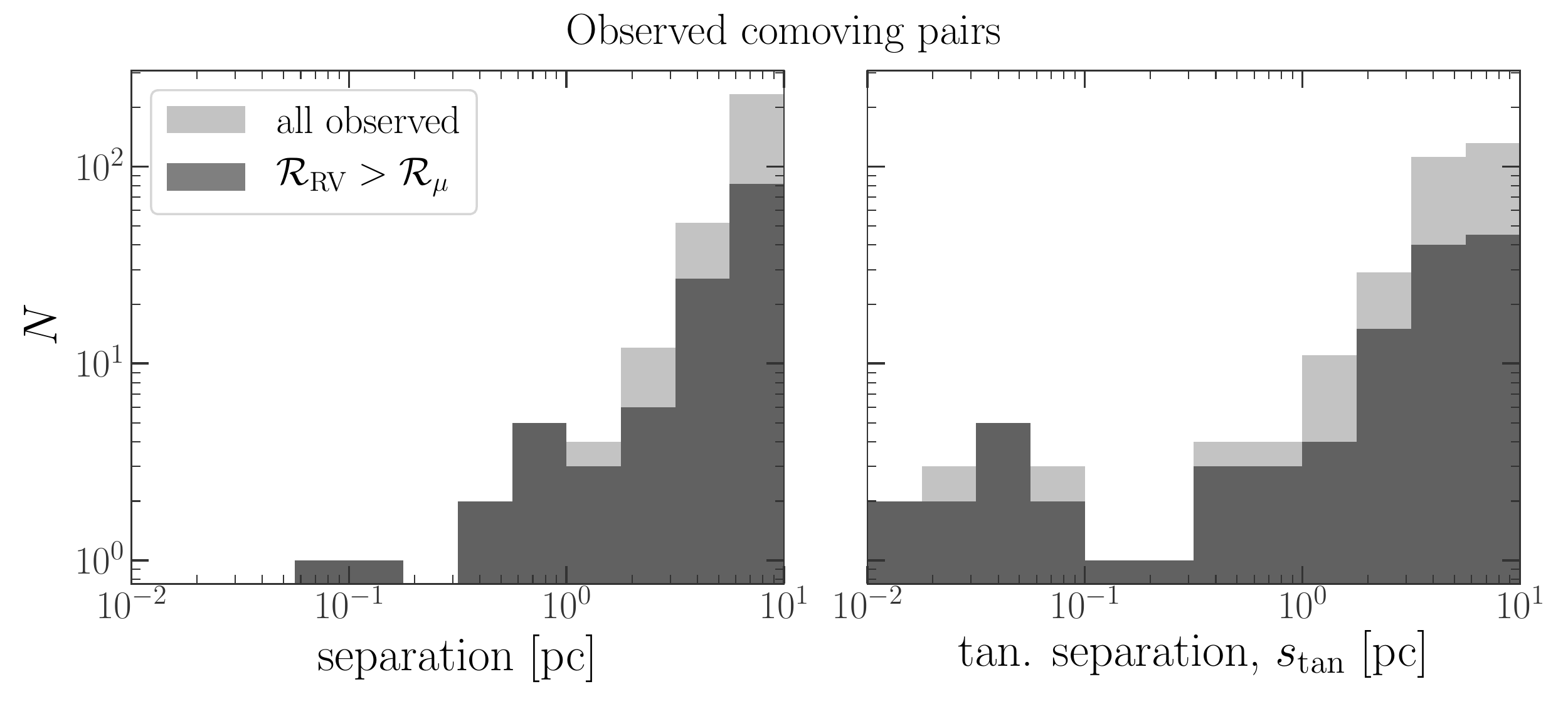}
  \end{center}
  \caption{%
    \emph{Left}: Distribution of point estimates of the 3D spatial separation of
    the stars in all observed comoving star pairs (gray), and only those
    confirmed as comoving (black).
    Note that the false-positive rate is $\approx 60\%$ at large separations
    ($>1~\pc$), but there are still a significant number of comoving pairs at
    these very wide separations.
    \emph{Right}: Distribution of tangential separations for all comoving star
    pairs (gray) and only those identified as comoving with 3D velocity
    information (black).
    Noise in the parallax measurements makes the 3D separation distribution
    appear roughly uniform.
    Even the distribution in tangential separation appears to be approximately
    linear for large separations ($\gtrsim 0.1~\pc$).
    \label{fig:separation}}
\end{figure}

\figurename~\ref{fig:separation} shows histograms of the spatial separation and
sky separation for all pairs; background, gray histogram shows all observed
pairs, and foreground, black histogram shows all pairs for which $\llrnew >
\llrold$.
In this and subsequent figures, we use the phrase ``tangential separation'' (and
symbol $s_{\rm tan}$) to refer to the chord length separation between the two
pairs assuming the two stars are at the same distance (the smaller distance of
the two, $d_{\rm min}$):
\begin{equation}
    s_{\rm sky} = 2\, d_{\rm min} \, \sin{\left( \frac{1}{2}\theta \right)}
\end{equation}
where $\theta$ is the angular extent of the pair.

At sky separations $\gtrsim 1~\pc$, $\approx 40\%$ of the comoving pairs are
confirmed to be comoving; interestingly, the separation distribution of the
genuine pairs appears to be uniform or possibly growing in number with
increasing separation.
This strongly suggests the existence of a population of unbound wide binaries or
dissolved open clusters, which maintain approximate spatial coherence for long
times because of the small relative velocities at the time of disruption of the
constituent stars (e.g., \citealt{Jiang:2010}).
The kinematic properties of these unbound binaries and associations contain
information about the tidal field and fluctuations in the mass distribution of
the Milky Way.
With \gaia\ \DR{2}, we will be able to study the statistical properties of these
widely-separated but comoving stars at even larger separations ($>10~\pc$).

\begin{figure}[p]
  \begin{center}
    \includegraphics[width=\linewidth]{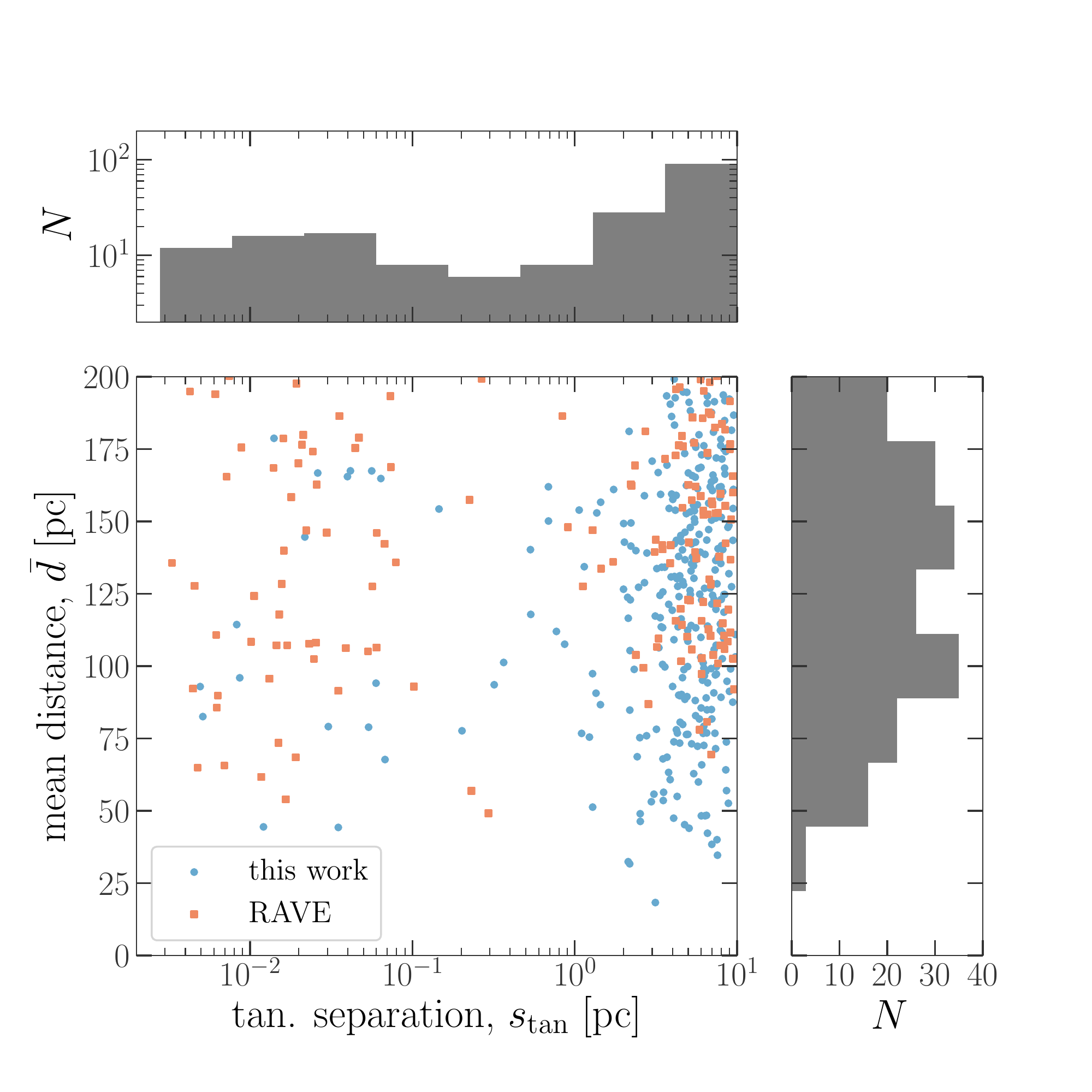}
  \end{center}
  \caption{%
    \emph{Top}: Distribution of tangential separation for all confirmed comoving
      star pairs, either from the radial velocity follow-up presented in this
      work or from RAVE \DR{5} radial velocities.
      At separations $\lesssim 10^{-1}~\pc$, the distribution (in log-log)
      appears uniform, consistent with past measurements of the wide binary
      separation distribution.
      At separations between $\approx 10^{-1}$--$10^{0}~\pc$, there appears to
      be an underabundance of comoving pairs, but the number of comoving pairs
      then begins to increase again out to our sample separation limit of
      $10~\pc$.
    \emph{Middle}: Tangential separation plotted against mean distance for each
      comoving pair identified from this work (circles, blue) or from RAVE
      overlap (squares, orange).
    \emph{Right}: Distance distribution for all confirmed comoving pairs.
      Note that at distances $\gtrsim 150~\pc$, our detection efficiency appears
      to decline.
    \label{fig:separation-RAVE}}
\end{figure}

An additional 202 candidate comoving pairs (\citealt{Oh:2017}) have
previously-measured radial velocities (for each star) reported in \DR{5} of the
Radial Velocity Experiment (RAVE; \citealt{Kunder:2017}).
Using the same procedure described in \sectionname~\ref{sec:catalog}, we
identify confirmed comoving pairs using the RAVE velocities for all pairs within
$200~\pc$ and find a similar false-positive rate at large separations ($\approx
38\%$ are confirmed).
Of the \nraveoverlap\ RAVE-overlap candidate comoving pairs (within $200~\pc$)
that meet the above criteria, we find \nravecomoving\ confirmed comoving pairs.
We plot the confirmed comoving pairs identified through radial velocity
follow-up in this work along with the RAVE-confirmed pairs in
\figurename~\ref{fig:separation-RAVE}.
Top histogram shows the tangential separation distribution for the joint
sample, scatter plot shows log-separation and the pair distances, and right
histogram shows the distance distribution of the confirmed pairs.
At larger distances ($\gtrsim 150~\pc$) it appears that the detection efficiency
of our method drops, as expected given the hard cut in log-FML-ratio applied to
define the candidate sample.
Reassuringly, there do not appear to be any trends of the separation
distribution with distance.

\begin{figure}[htbp]
  \begin{center}
    \includegraphics[width=\linewidth]{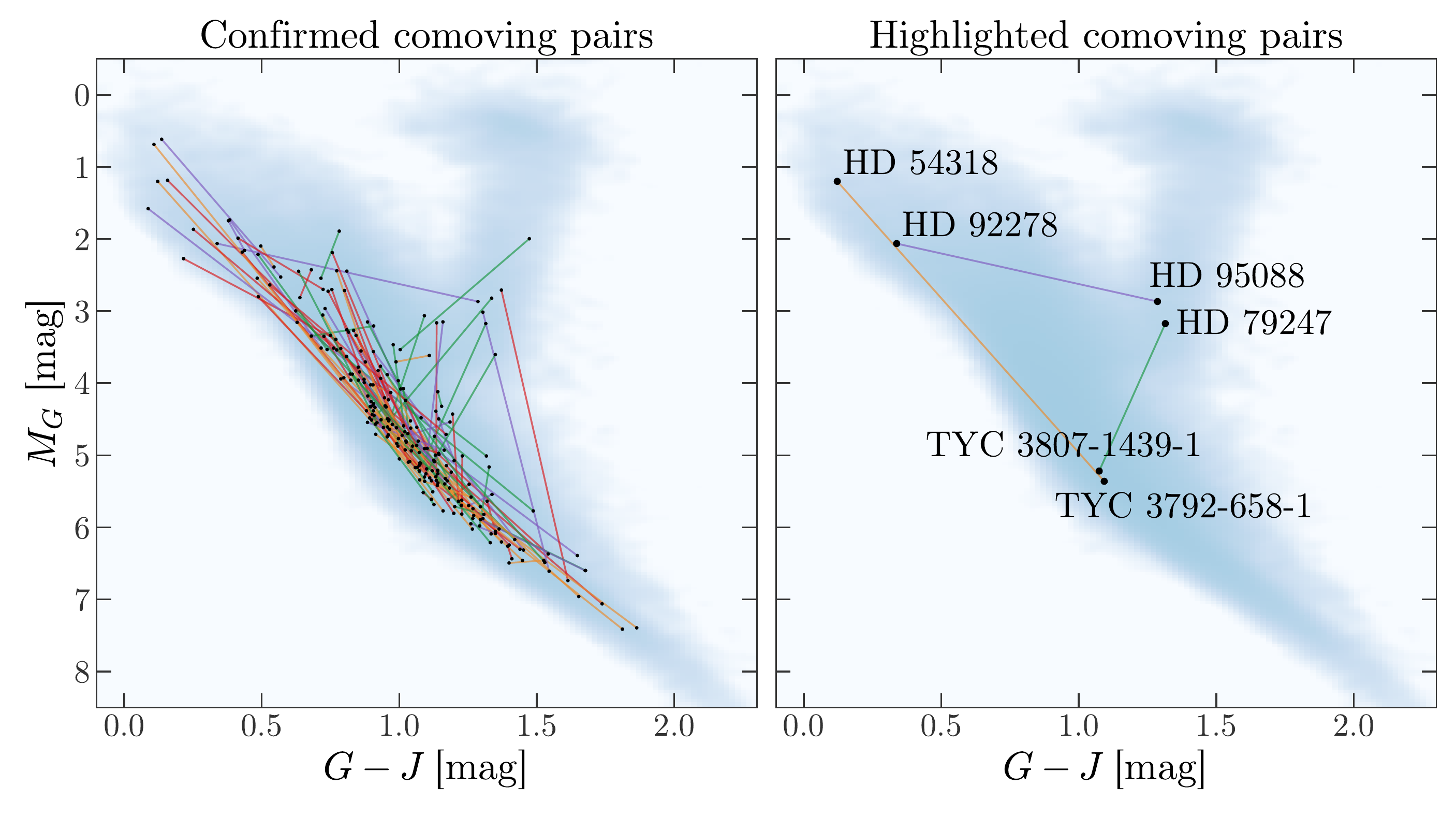}
  \end{center}
  \caption{%
    {\it Left}: Color-magnitude diagram of confirmed comoving star pairs plotted
    over the density of all \tgas\ sources within 200 pc; similar to
    \figurename~\ref{fig:sample-cmd}, but for only the confirmed comoving pairs.
    Several interesting pairs that include main-sequence and giant component
    stars are highlighted in the text (\sectionname~\ref{sec:interesting-pairs})
    and in the right panel.
    {\it Right}: Same as left panel, but only three highlighted comoving star
    pairs are shown.
    \label{fig:genuine-highlighted-cmd}}
\end{figure}

\figurename~\ref{fig:genuine-highlighted-cmd} (left) shows the color-magnitude
diagram for all confirmed comoving pairs.
As would be expected if all of the pairs are coeval, the majority of pairs lie
along isochrones in color-magnitude space, with the exception of a few visible
outliers.
For example, there appear to be at least a few main-sequence--giant-branch
pairs, and a few pairs that appear to cross the main-sequence.
We highlight a few of these pairs in the next section
(\sectionname~\ref{sec:interesting-pairs}).

\clearpage

\subsection{A few highlighted comoving pairs}\label{sec:interesting-pairs}

\figurename~\ref{fig:genuine-highlighted-cmd} (right) shows a few highlighted
pairs selected from the color-magnitude diagram of confirmed comoving stars.
We compile \project{Tycho-2} ($B$, $V$; \citealt{Hog:2000}), \tmass\
($J$, $H$, $K_{\rm s}$; \citealt{Skrutskie:2006}), and \acronym{WISE} ($W_1$,
$W_2$, $W_3$; \citealt{Wright:2010}) photometry for each star in the pairs below
and fit isochrones to the stars individually to measure their stellar
parameters.
We use the \package{isochrones} \python\ package (\citealt{Morton:2015}) to fit
the photometric and parallax information (and uncertainties) for each star to
infer the \logg, \teff, and stellar mass; these values are reported in
\tablename~\ref{tbl:highlighted-pairs}.
In brief, we use the model grids from the Dartmouth Stellar Evolution Database
(\citealt{Dotter:2008}) and the built-in interpolation in \package{isochrones}
to compute the posterior probability of a given set of stellar parameters given
the data (photometry and parallax).
We generate samples (in stellar parameters) from this posterior pdf using the
ensemble MCMC sampler implemented in \package{emcee}
(\citealt{Foreman-Mackey:2013}) using 128 walkers.
We run the sampler for 256 initial steps from random initial positions, compute
the median position of the walkers in parameter-space and re-initialize the
walkers in a small Gaussian ball around this median position to run for another
1024 burn-in steps.
We then discard all but the last position and re-run for a final 8192 steps,
saving only every 128th step.
The stellar parameter values listed below are computed by taking the median
posterior sample, and the uncertainties are estimated by computing the median
absolute deviation, ${\rm MAD}$, and converting to a standard deviation under
the assumption of Gaussianity ($\sigma \approx 1.5 \times {\rm MAD}$).

\begin{table}[ht]
  \begin{center}
    \begin{tabular}{ c | c | c | c | c }
      \toprule
        Star name & \logg\ & \teff\ [K] & mass [$M_\odot$] &
          tangential separation [pc]  \\
        \toprule
        HD 95088 & $3.7 \pm 0.02$ & $5303 \pm 74$ & $1.490 \pm 0.002$ &
          \multirow{2}{*}{$8.6 \pm 0.8$}\\
        HD 92278 & $4.2 \pm 0.05$ & $8519 \pm 300$ & $1.8 \pm 0.1$ &\\
        \midrule
        HD 79247 & $3.61 \pm 0.02$ & $5075 \pm 60$ & $1.150 \pm 0.004$ &
          \multirow{2}{*}{$6.8 \pm 0.3$}\\
        TYC 3807-1439-1 & $4.50 \pm 0.04$ & $5409 \pm 80$ & $0.88 \pm 0.05$ &\\
        \midrule
        HD 54318 & $4.15 \pm 0.04$ & $10030 \pm 345$ & $2.4 \pm 0.1$ &
          \multirow{2}{*}{$8.4 \pm 1.1$}\\
        TYC 3792-658-1 & $4.54 \pm 0.03$ & $5462 \pm 139$ & $0.89 \pm 0.06$ &\\
      \bottomrule
    \end{tabular}
    \caption{
      Stellar parameters inferred using isochrone-fitting for the three pairs
      highlighted in \sectionname~\ref{sec:interesting-pairs}.
      \label{tbl:highlighted-pairs}
    }
  \end{center}
\end{table}

\subsubsection{HD 95088 -- HD 92278}

From their surface gravities and temperatures, the component stars of this pair
appear to be an A-type dwarf (HD 92278) and a slightly less massive G-type
subgiant (HD 95088).
This pair is separated by $4.5^\circ$ on the sky (corresponding to $\approx
8.6~\pc$ in tangential separation, see \tablename~\ref{tbl:highlighted-pairs})
and have consistent distances within their uncertainties of $\approx 110 \pm
5~\pc$.
At first glance, it seems unlikely that the two stars could be coeval, as the
more massive A star has not begun its post-main-sequence evolution.
One possible explanation is that this pair is a false-positive in three
dimensions, in which case they are unassociated but just appear to be comoving
with small relative velocity at present day.

Other speculative explanations for this pair stem from the possibility that it
is the outcome of an interaction between even more stars.
For example, an interaction between two wide binaries could result in an
exchange of members, a hardening of one resulting binary, and a widening of the
other binary (see, e.g., \citealt{Leigh:2016}).
Or, the initial system could have been an unstable three-body system that
resulted in a tight inner binary and a wide outer binary;
if the inner binary were close enough, it could have merged and formed a blue
straggler star (e.g., \citealt{Naoz:2014}), leaving the wide binary untouched.

High-resolution spectroscopy of these stars --- and precise measurements of the
stellar parameters and chemical abundances --- may help to distinguish these
scenarios.

\subsubsection{HD 79247 -- TYC 3807-1439-1}

The components of this pair appear to be a K-type subgiant (HD 79247) and a
G-type dwarf (TYC 3807-1439-1).
This pair is separated by $4^\circ$ on the sky or $6.8~\pc$ in tangential
separation and have an apparent radial separation of $\approx 3~\pc$,
corresponding to a 3D separation of $\approx 9.8~\pc$ at a distance of $\approx
100~\pc$.
Coeval stars at different phases of stellar evolution are extremely valuable for
calibrating and testing stellar evolution models (e.g., \citealt{Torres:2013}).
With a larger sample of confirmed comoving pairs, identified pairs that contain
both an evolved star and a main sequence star --- such as this pair --- will be
of great interest.

\subsubsection{HD 54318 -- TYC 3792-658-1}

The components of this pair appear to be an A- or B-type dwarf (HD 54318) and a
G-type dwarf (TYC 3792-658-1), the largest magnitude difference between all of
the confirmed comoving pairs.
This pair is separated by $2.5^\circ$ on the sky or $8.5~\pc$ in tangential
separation and have a consistent distance of $\approx 190~\pc$ within their
uncertainties.
The mass ratio distribution of comoving star pairs will help constrain star
formation models (e.g., \citealt{Raghavan:2010}).

\section{Discussion} \label{sec:discussion}


The results presented in this work demonstrate that even at separations larger
than a parsec, a substantial fraction of comoving pair candidates selected from
astrometry alone are confirmed with three-dimensional velocity information.
This is exciting because it confirms the existence of what is likely a
population of nearly-coherent but unbound coeval stars.
It is also encouraging for the prospect of selecting large samples of comoving
pairs purely with astrometric data from future data releases of the \gaia\
mission (radial velocity measurements will only be released for bright stars, $G
< 12$).
However, in this work, we do not claim to report the true number or density of
large-separation comoving pairs.
To model the abundance and formation of comoving pairs, it will be crucial to
select larger samples of pairs without hard cuts on separation, tangential
velocity, signal-to-noise, or likelihood ratio, as were done with the input
sample used in this work.
It will also be important to quantify the contamination rate of (3D) comoving
pair lookalikes, especially at large separations.

\begin{figure}[htbp]
  \begin{center}
    \includegraphics[width=0.9\linewidth]{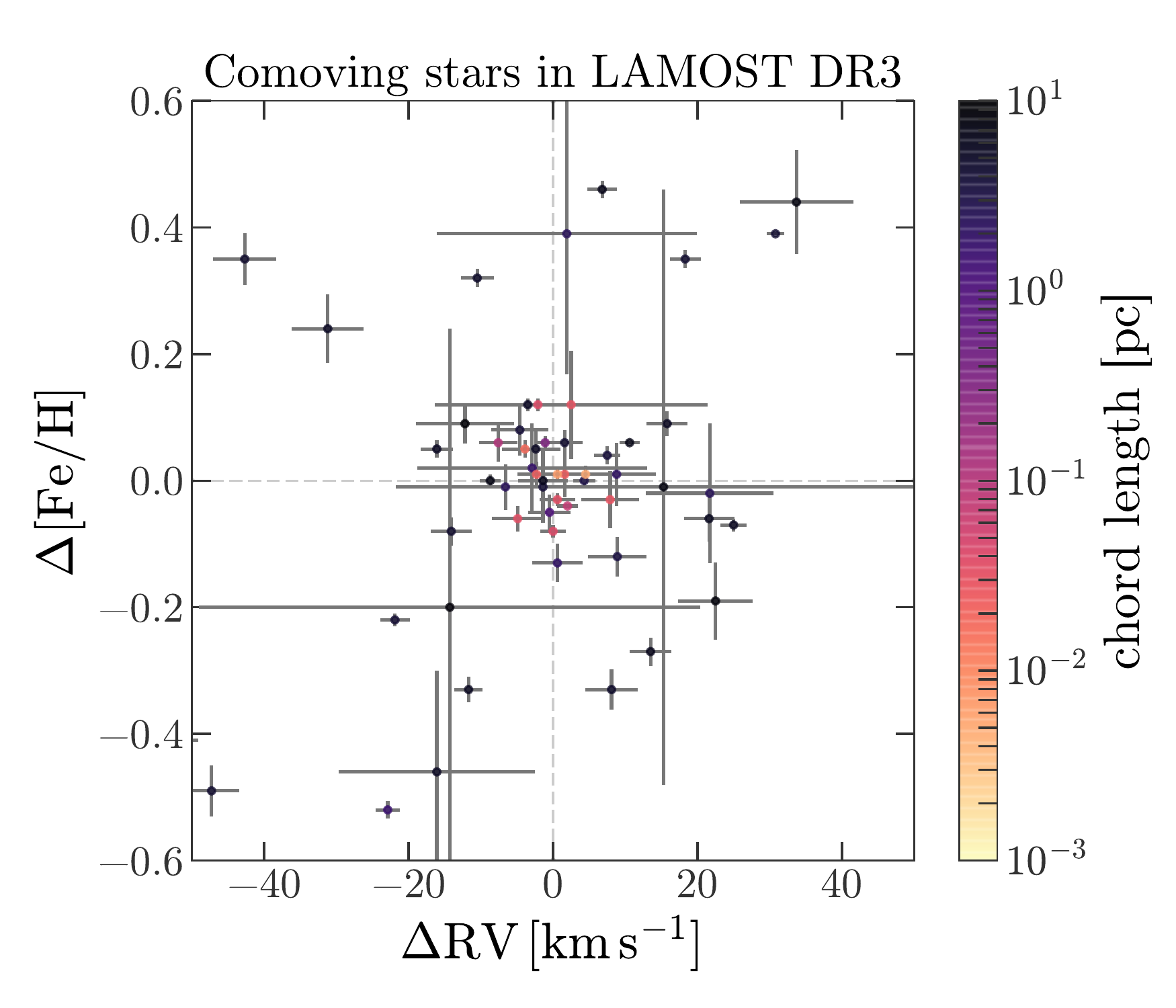}
  \end{center}
  \caption{%
    Differences in metallicity and radial velocity for candidate comoving pair
    member stars from the \acronym{LAMOST} survey.
    Points represent each pair, colored by the tangential separation, and
    error bars represent the uncertainties taken from the \acronym{LAMOST}
    \DR{3} catalog.
    There is a clear over-density of pairs with near-zero differences in
    metallicity and radial velocity even at large ($>1~\pc$) separations, adding
    even more confidence to the existence of comoving star pairs with these
    extreme separations.
    \label{fig:lamost}}
\end{figure}

We do not try to estimate a contamination rate in this work, but instead use
metallicities and radial velocities from 61 candidate comoving pairs that are in
\DR{3} of the \acronym{LAMOST} survey (\citealt{Deng:2012}) as an additional
validation test.
\figurename~\ref{fig:lamost} shows these pairs plotted as the difference in
radial velocity and difference in iron abundance with points colored by the
tangential separation of each pair.
There is a clear over-density of pairs with abundance and radial-velocity
differences close to zero.
From this figure it is also clear that the false positives are predominantly the
largest separation pairs (i.e. are the darkest markers), but a number of
$>1~\pc$ pairs appear to have small radial-velocity differences and small
differences in iron abundances, making them even more likely to be genuine
pairs.

Going forward, definitive confirmation of comoving star pairs may require
precise measurements of chemical abundances combined with precise radial
velocities, as demonstrated with the \acronym{LAMOST} data.
However, we would also like to \emph{test} assumptions about chemical
homogeneity, rather than simply assuming that all true comoving pairs must have
identical abundances.
For example, the outcomes of planet formation may lead to large or appreciable
abundance differences between coeval stars (e.g., \citealt{Pinsonneault:2001},
Oh et al., submitted).
We therefore caution that---to study the chemical properties of comoving pairs
and star formation in general---the comoving pairs should be selected from their
kinematics alone.

\section{Conclusions}

The information contained in a large sample of widely-separated, comoving star
pairs spans from the dynamics of stars and dark matter in the Galaxy, the
fundamentals of star formation, and the stability of planetary systems.
Through our own low-resolution spectroscopic follow-up and a cross-match with
the RAVE \DR{5} catalog, we have identified \ncomovingtotal\ comoving star pairs
by combining these radial velocity measurements with astrometry from the \tgas\
catalog.
From a cross-match to the \acronym{LAMOST} survey, we find that candidate
comoving star pairs with small differences in radial velocities also have small
differences in iron abundance, even at large separations ($>1~\pc$).
The separation distribution of the spectroscopically-confirmed comoving star
pairs presented shows an abundance of comoving pairs at separations between
$1$--$10~\pc$ (\figurename~\ref{fig:separation-RAVE}), the imposed limit in our
initial search (\citealt{Oh:2017}).
This result is further tentative evidence (see also \citealt{Shaya:2011}) for
the existence of a population of unbound comoving stars that represent the
ultimate fate of wide binaries and moving groups that slowly dissociate due to
weak perturbations from stars (\citealt{Jiang:2010}) and other massive
perturbers in the Galaxy.
From this work it appears that modeling the population of wide binaries or
comoving stars with a monotonically decreasing separation distribution---as is commonly done---too simple to also represent the population of unbound systems.
With future data releases from the \gaia\ mission combined with large
spectroscopic surveys, the anticipated catalogs of confirmed comoving star pairs
will offer a new ``stereo'' view on the evolution and state of our Galaxy.

\acknowledgements

It is a pleasure to thank
Ruth Angus (Columbia),
John Brewer (Yale),
Jason Curtis (Columbia),
Keith Hawkins (Columbia),
David W. Hogg (NYU),
Nathan Leigh (AMNH),
Melissa Ness (MPIA),
Dave Zurek (AMNH),
for useful discussions and advice.
We also thank all attendees of the ``Stars'' group meeting at the Center for
Computational Astrophysics at the Flatiron Institute.
The Flatiron Institute is supported by the Simons Foundation.

We thank James Davenport (WWU) for releasing his open source spectroscopic
reduction code, \package{PyDIS}, which served as inspiration for the pipeline
used in this work.
This work is based on observations obtained at the MDM Observatory, operated by
Dartmouth College, Columbia University, Ohio State University, Ohio University,
and the University of Michigan.
This work has made use of data from the European Space Agency (ESA)
mission {\it Gaia} (\url{https://www.cosmos.esa.int/gaia}), processed by
the {\it Gaia} Data Processing and Analysis Consortium (DPAC,
\url{https://www.cosmos.esa.int/web/gaia/dpac/consortium}). Funding
for the DPAC has been provided by national institutions, in particular
the institutions participating in the {\it Gaia} Multilateral Agreement.
This research has made use of the SIMBAD database, operated at CDS, Strasbourg,
France.
This project was developed in part at the 2017 Heidelberg Gaia Sprint, hosted by
the Max-Planck-Institut für Astronomie, Heidelberg.

Guoshoujing Telescope (the Large Sky Area Multi-Object Fiber Spectroscopic
Telescope LAMOST) is a National Major Scientific Project built by the Chinese
Academy of Sciences. Funding for the project has been provided by the National
Development and Reform Commission. LAMOST is operated and managed by the
National Astronomical Observatories, Chinese Academy of Sciences.

\software{
The code used in this project is available from
\url{https://github.com/adrn/GaiaPairsFollowup} under the MIT open-source
software license.
This research utilized the following open-source \python\ packages:
    \package{Astropy} (\citealt{Astropy-Collaboration:2013}),
    \package{astroquery} (\citealt{Ginsburg:2016}),
    \package{ccdproc} (\citealt{Craig:2015}),
    \package{celerite} (\citealt{Foreman-Mackey:2017}),
    \package{corner} (\citealt{Foreman-Mackey:2016}),
    \package{emcee} (\citealt{Foreman-Mackey:2013ascl}),
    \package{IPython} (\citealt{Perez:2007}),
    \package{matplotlib} (\citealt{Hunter:2007}),
    \package{numpy} (\citealt{Van-der-Walt:2011}),
    \package{scipy} (\url{https://www.scipy.org/}),
    \package{sqlalchemy} (\url{https://www.sqlalchemy.org/}).
This work additionally used the Gaia science archive
(\url{https://gea.esac.esa.int/archive/}), and the SIMBAD database
(\citealt{Wenger:2000}).
}

\facility{MDM: Hiltner (Modspec)}

\appendix

\section{1D spectrum extraction} \label{sec:1d-ext}

For each source or comparison lamp observation, the two-dimensional image is
bias corrected, flat-field corrected, and trimmed using standard routines
implemented in \package{ccdproc} (\citealt{Craig:2015}).
The 2D spectra are oriented such that the spatial direction is along a row of
pixels, and the dispersion direction is along a column of pixels.
For comparison lamp observations, the central 100 pixel columns are extracted
and summed along the spatial axis, weighted by the inverse-variances of each
pixel.
Source spectral traces were always placed in this region during observations,
and the curvature of the wavelength solution over this region is negligible.
For source observations, 1D spectra are extracted using an empirical
estimate of the line spread function (LSF).
The LSF is assumed to be a Voigt profile with parameters $A$, amplitude; $x_0$,
profile centroid; $\sigma_G$, Gaussian root-variance; $\gamma_L$, Lorentzian
half-width at half maximum, and the background is modeled using a linear
polynomial with parameters $\alpha_1$ and $\alpha_2$, the slope and intercept
at the profile centroid, respectively.
The maximum-likelihood LSF + background model parameters are estimated from
each pixel row using \lbfgsb\ optimization (\citealt{Zhu:1994}), implemented in
the \package{scipy} package.
We first determine the LSF width parameters, $\sigma_G$ and $\gamma_L$ from fits
to 16 pixel rows evenly spaced between indices 750 and 850 (i.e. around the
effective center of the dispersion axis).
From these 16 fits, the median profile width parameters, $\sigma_G$ and
$\gamma_L$, are then taken to be the LSF width parameters.
Then, at each of 1600 rows in each source spectrum image, the LSF amplitude and
centroid, and background slope and intercept at line centroid, are then fit
using the same procedure as above but with the width parameters fixed.
The LSF amplitudes are stored as the source fluxes, and the background model
intercepts are stored as the background (sky) fluxes.
Occasionally, nearby sources fall in the slit; these are masked by hand in the
2D images and do not bias the LSF extraction fits.

\section{Wavelength calibration}

We start the wavelength calibration by interactively identifying the 12
strongest, least blended emission lines in one of the extracted 1D comparison
lamp spectra by comparing to standard HgNe line lists.\footnote{\url{http:
//physics.nist.gov/PhysRefData/ASD/lines_form.html}}
Once a rough mapping from pixel to wavelength is determined, we fit
8 pixel-wide segments around each identified emission line
for the individual line centroids in each comparison lamp spectrum.
We use a pixel-convolved Gaussian profile for each emission line with
parameters: amplitude, $A$; centroid, $x_0$; and root-variance, $\sigma_G$.
We find that nearby blended or low-amplitude lines (i.e. part of the
``background'') sometimes bias the line centroid determination when using a
simple (i.e. low-order polynomial) background model.
We therefore model the background flux around each comparison lamp emission
line using a constant offset + Gaussian process (GP) in order to capture any
unmasked structure from nearby sources.
For the GP model, we use a Mat\'ern 3/2 kernel function
(\citealt{Matern:1986,Rasmussen:2005}) parametrized by amplitude parameter
$\sigma_{3/2}$ and correlation scale $\rho_{3/2}$.
We use \package{celerite} (\citealt{Foreman-Mackey:2017}) to evaluate the full
likelihood for this model, and maximize the (log-)likelihood again using
\lbfgsb\ optimization.
We optimize over the logarithm of any scale parameters (i.e. $A$, $\sigma_G$,
$\sigma_{3/2}$, and $\rho_{3/2}$).
Once we determine the pixel centroids for all identified emission lines for a
given night's comparison lamp spectrum, we then fit an interpolating function
to the pixel-centroid, wavelength values for each emission line.

\begin{figure}[htbp]
  \begin{center}
    \includegraphics[width=0.8\linewidth]{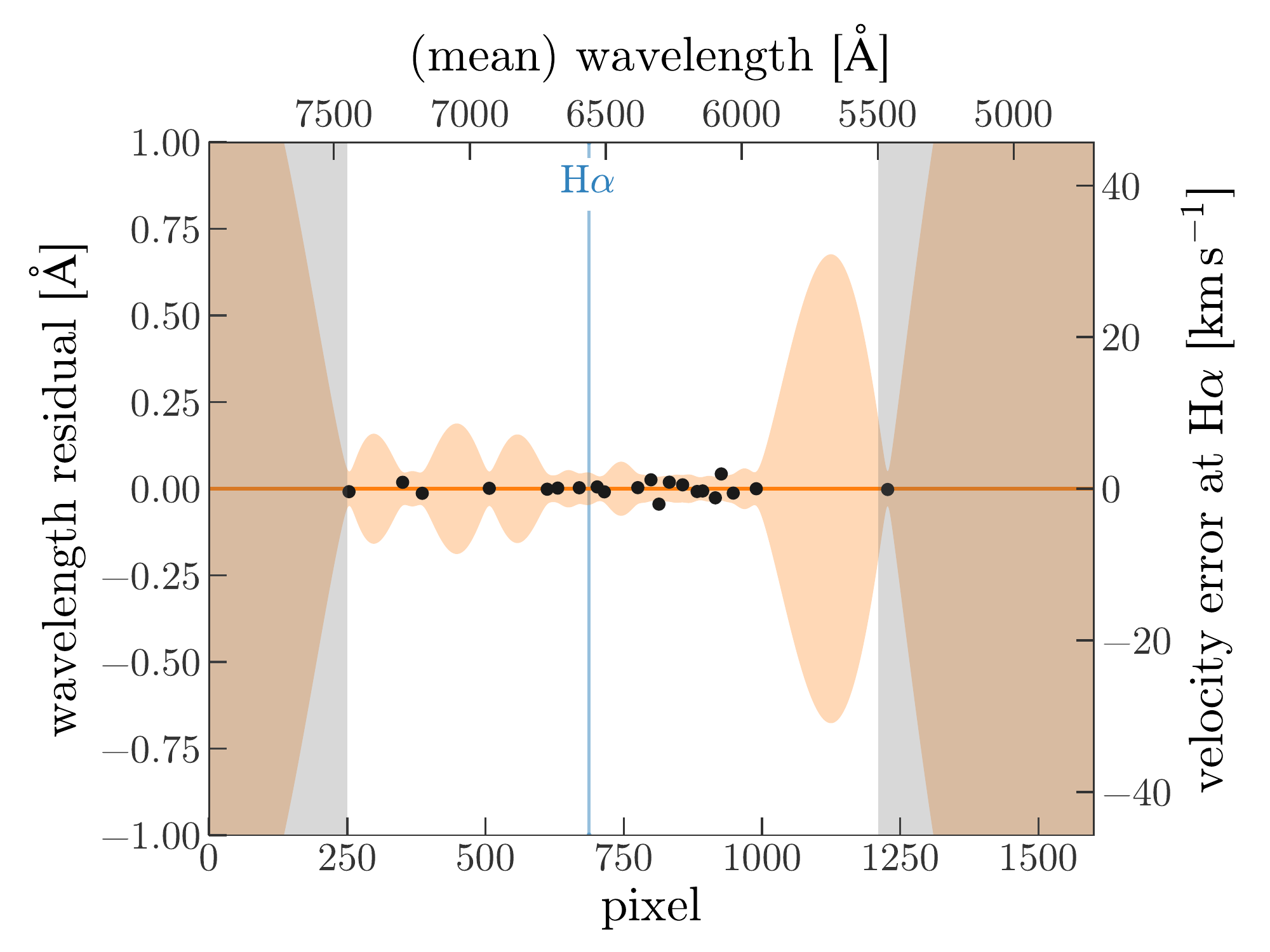}
  \end{center}
  \caption{%
    A visualization of the pixel-to-wavelength model determined from a single
    wavelength comparison lamp spectrum.
    Black markers show the fit pixel centroids for known Hg or Ne emission
    lines and the residual wavelength value between the linear + GP model and
    the true emission line centroid.
    The non-shaded region shows the section of spectrum where the comparison
    lamp spectrum has sufficient line density to determine the wavelength
    solution; shaded regions are masked from the resulting spectra.
    Orange envelope shows the root-variance of the Gaussian process component
    of the model, which can be interpreted as inherent uncertainties on the
    wavelength calibration as a result of the model flexibility.
    \label{fig:wavelength-GP}}
\end{figure}

Typically, high-order standard polynomials or Chebyshev polynomials are used to
fit for the wavelength dispersion function.
We found that using such functions can lead to vastly incorrect wavelength
values near either end of the pixel grid where the polynomial function is
un- or weakly-constrained, and can easily over-fit to poorly centroided emission
lines.
We instead use a combined linear polynomial + GP to model the wavelength
dispersion to flexibly account for nonlinear behavior in the pixel to wavelength
mapping.
We again use a Mat\'ern 3/2 kernel, use \lbfgsb\ optimization, and convert any
scale parameter to log-space before maximizing the log-likelihood.
\figurename~\ref{fig:wavelength-GP} shows an example of the residuals from one
such fit; black points show the (pixel-centroid, wavelength) pairs with the
best-fit model subtracted.
The envelope around zero (shaded region, orange) shows the variance of the GP
model projected into the data space; when the GP variance is large, the
pixel-to-wavelength mapping is uncertain.
We therefore only use the pixel range from 250--1210, corresponding to a
wavelength range of $\approx 5500$--$7400~{\rm \AA}$; gray shaded region shows
the excluded pixel/wavelength ranges.

\begin{figure}[htbp]
  \begin{center}
    \includegraphics[width=0.6\linewidth]{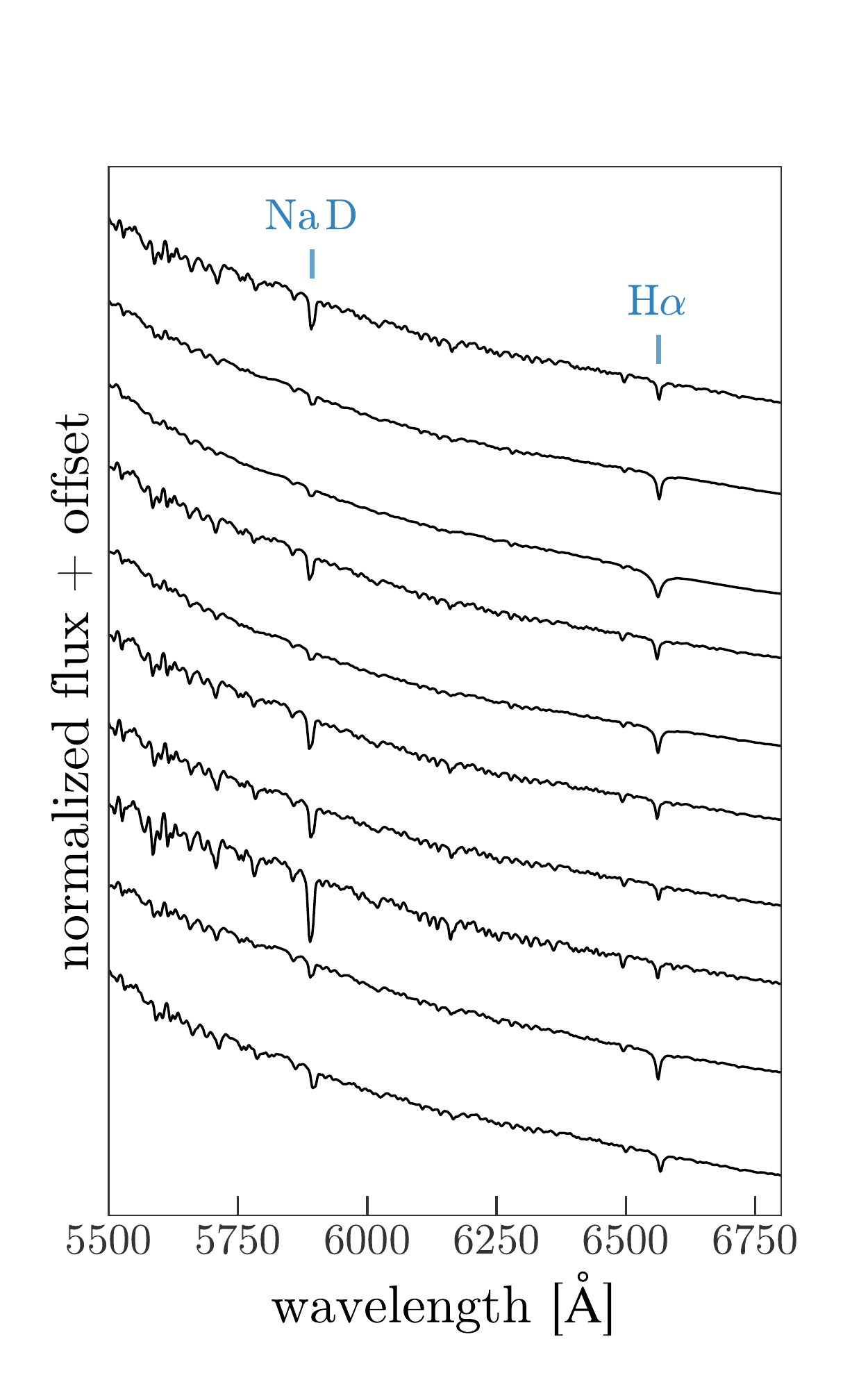}
  \end{center}
  \caption{%
    A section of 10 randomly selected target spectra from this work.
    Spectral fluxes are normalized to 1 at $\lambda = 5500~{\rm \AA}$ and the
    offsets are chosen to separate the spectra for visualization.
    The spectra are smoothed with a Gaussian filter with root-variance $\sigma
    = 2~{\rm \AA}$.
    \label{fig:sample-spectra}}
\end{figure}

\figurename~\ref{fig:sample-spectra} shows a random sample of 10
wavelength-calibrated source spectra of comoving star pair members normalized
to 1 at $\lambda = 5500~{\rm \AA}$ and shifted by an arbitrary offset for
visualization.
As suggested by the features (the number and depths of absorption features) in
these 10 spectra, there is a mix of metallicities and spectral types amongst the
observed targets.
Note that the spectra are not flux-calibrated, as we only care about measuring
radial velocities from this sample.

The wavelength calibration procedure described above produces a nightly model
for the nonlinear dispersion of wavelength as a function of pixel value for
each extracted 1D spectrum.
However, because of flexure in the spectrograph, each observation will have an
additional offset of, typically, $\approx 0.5$--$2~{\rm \AA}$, corresponding to
a significant velocity offset at \Ha\ of $\approx 25$--$100~\kms$.
As mentioned, comparison lamp spectra were not obtained at each telescope
pointing.
The absolute wavelength solution at each pointing can therefore be shifted
(in pixels) relative to the nightly comparison lamp spectrum because of flexure
in the instrument.
We have found that this offset depends primarily on the hour angle of the
observation, likely because of the physical orientation of the spectrograph and
CCD as mounted on the telescope.
To correct each source spectrum for this offset, we measure the centroids of
the two prominent [OI] sky emission lines near $5577~{\rm \AA}$ and $6300~{\rm
\AA}$ in the corresponding background spectrum derived from the spectral
extraction (described above).
We use a pixel-convolved Voigt profile as the emission line model (same
parameters as above) and a Gaussian process with a Mat\'ern 3/2 kernel function
for the background spectrum (same parameters as above).
We again use \lbfgsb\ optimization and convert any scale parameter to log-space
before maximizing the log-likelihood of this model to derive the line centroid
in wavelength units.
We compute wavelength offsets from true (air) values of the line centroids,
$5577.3387~{\rm \AA}$ and $6300.304~{\rm \AA}$ respectively,\footnote{\url{http:
//physics.nist.gov/PhysRefData/ASD/lines_form.html}} then convert these
wavelength offsets to pixel offsets.
We take the mean of the pixel offsets and apply the derived shift to the source
spectrum before re-computing the wavelength grid using the nightly wavelength
solution.
When one or both of the two sky lines is not present or the centroid fit fails,
the spectrum is flagged as having a suspect wavelength calibration.
In practice, most such sources are calibration targets, which are more likely to
have very short exposure times and thus weak night-sky lines.

\begin{table}[ht]
  \begin{center}
    \begin{tabular}{ c | c | l }
      \toprule
        Parameter & Prior & Description \\\toprule
        $\ln A$ & $\mathcal{U}(2, 16)$ & Logarithm of line amplitude
          [${\rm photons}$]\\
        $x_0$ & $\mathcal{U}(6547, 6579)$ & Line wavelength centroid
          [${\rm \AA}$]\\
        $\ln\sigma_G$ & $\mathcal{U}(-4, 2)$ & Logarithm of Gaussian standard
          deviation of line profile [${\rm \AA}$]\\
        $\ln\gamma_L$ & $\mathcal{U}(-4, 2)$ & Logarithm of Lorentzian HWHM of
          line profile\\
        $\alpha_1$ & $\mathcal{U}(0, 10^{16})$ & Background model offset at line
          centroid [${\rm photons}$]\\
        $\alpha_2$ & $\mathcal{U}(-10^8, 10^8)$ & Background model slope
          [${\rm photons}~{\rm \AA}^{-1}$]\\
        $\ln\sigma_{3/2}$ & $\mathcal{U}(-8, 8)$ & Logarithm of Gaussian process
          amplitude [${\rm photons}$]\\
        $\ln\rho_{3/2}$ & $\mathcal{N}(1, 1)$ & Logarithm of Gaussian process
          correlation scale [${\rm \AA}$]\\
      \bottomrule
    \end{tabular}
    \caption{
      Priors used for \Ha\ line profile model parameters.
      $\mathcal{U}(a, b)$ indicates a uniform prior between $a$ and $b$ and
      $\mathcal{N}(\mu, \sigma^2)$ indicates a normal distribution with mean
      $\mu$ and variance $\sigma^2$.
      Flux units are photons, i.e. the spectra are not flux-calibrated.
      \label{tbl:prior-bounds}
    }
  \end{center}
\end{table}

\section{Radial velocity measurement} \label{sec:rv-meas}

To derive radial velocities for each source, we then measure the (wavelength)
line centroid of \Ha\ in each corrected spectrum.
We model the absorption line using a pixel-convolved Voigt profile and use a
linear model + a Gaussian process for the background model.
We now use \lbfgsb\ optimization and the resulting maximum-likelihood parameters
to initialize a Markov Chain Monte Carlo (MCMC) sampling of the posterior
probability distribution (PDF) over the 8 parameters: $\ln |A|$, $x_0$,
$\ln\sigma_G$, $\ln\gamma_L$, $\alpha_1$, $\alpha_2$, $\ln\sigma_{3/2}$, and
$\ln\rho_{3/2}$ where the linear background model is defined as $f(\lambda) =
\alpha_1 + \alpha_2\,\lambda$.
The linear amplitude, $A$, is assumed to be negative (i.e. \Ha\ is assumed to be
an absorption feature) and is computed from the corresponding parameter as $A =
-\exp(\ln |A|)$.
We use uniform priors on all parameters except $\ln\rho_{3/2}$, for which we use
a Gaussian; prior parameters are specified in \tablename~\ref{tbl:prior-bounds}.
We use \package{emcee} (\citealt{Foreman-Mackey:2013}) with 64 walkers to
generate samples from the posterior PDF.
We initially run the sampler for 128 steps from the maximum likelihood parameter
values, then resample walker positions in a small ball around the median final
walker positions and run again for 512 steps; together, these two steps
constitute our burn-in phase.
From the final walker positions after burn-in, we run again for 1024 steps and
store these as our posterior samples.

\begin{figure}[!hp]
  \begin{center}
    \includegraphics[width=0.75\linewidth]{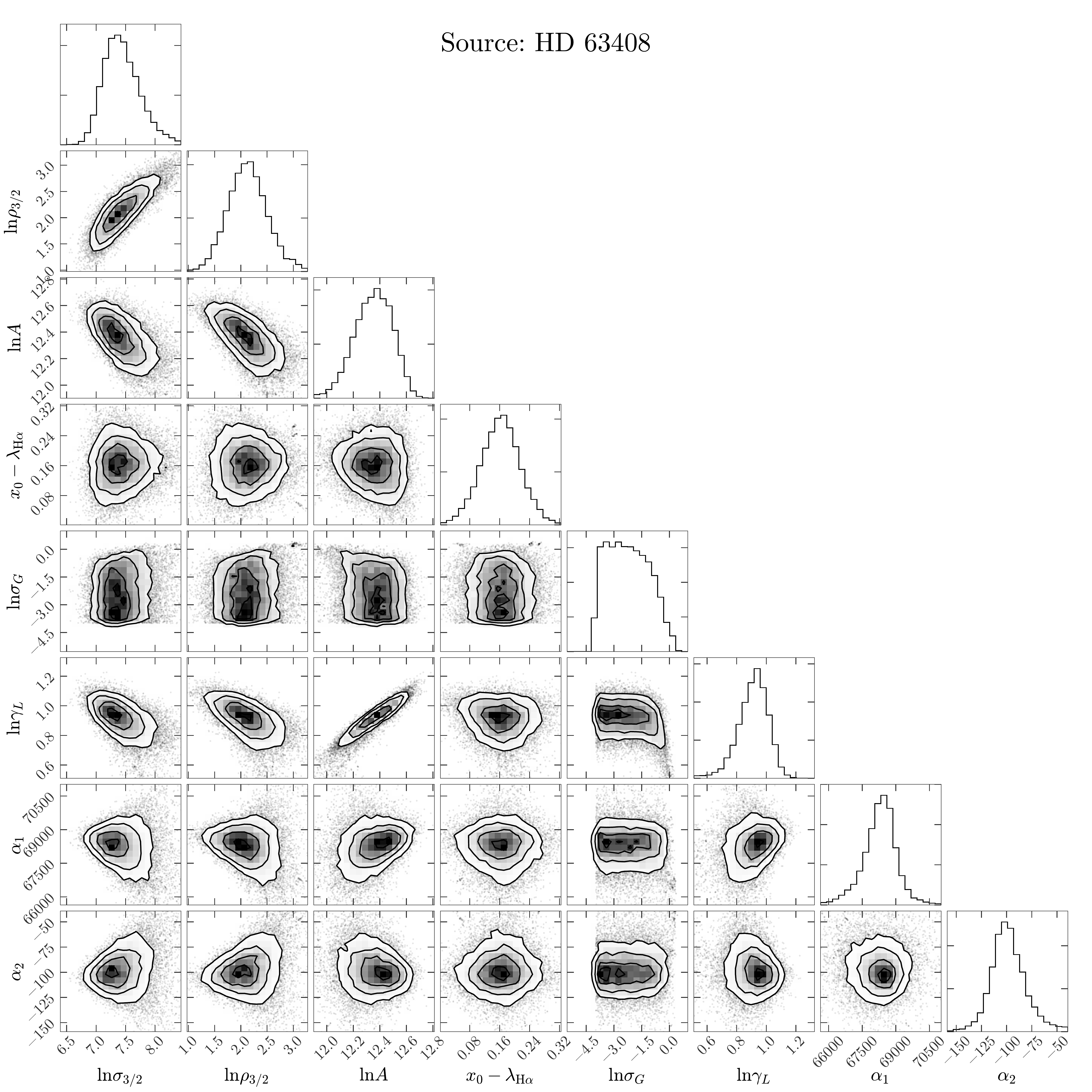}
  \end{center}
  \caption{%
    A corner plot showing all projections of posterior PDF samples for the \Ha\
    line model and the spectrum of the source HD 63408.
    Note the small uncertainty in the line centroid measurement, $x_0$, which
    corresponds to a $\approx 2~\kms$ velocity precision for this source.
    Parameters are described in \sectionname~\ref{sec:reduction}, and prior PDFs
    over the parameters are described in \tablename~\ref{tbl:prior-bounds}.
    \label{fig:Halpha-mcmc-corner}}
\end{figure}

\begin{figure}[htbp]
  \begin{center}
    \includegraphics[width=0.5\linewidth]{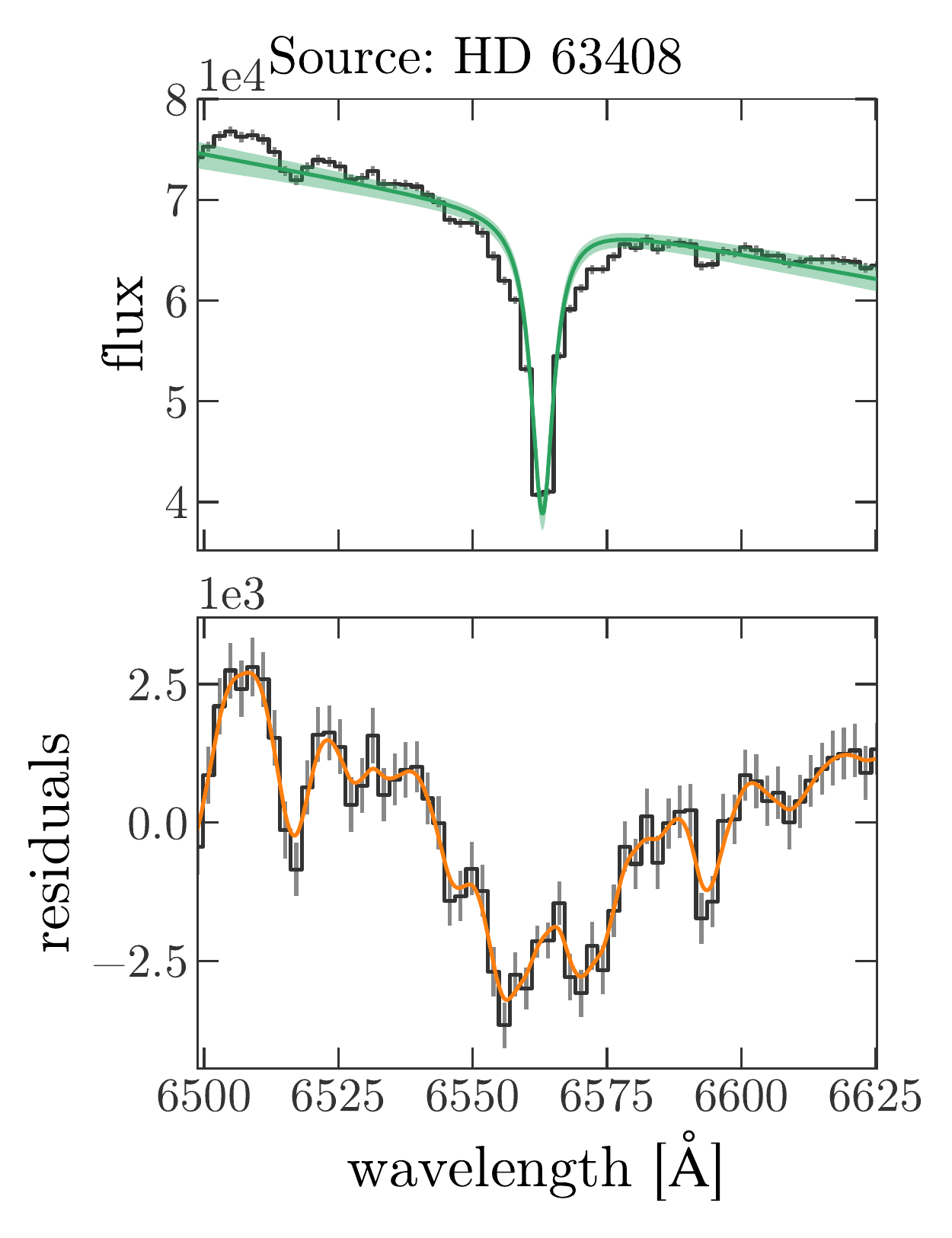}
  \end{center}
  \caption{%
    Visualization of the \Ha\ absorption line model and fits for the source HD
    63408 as a demonstration of the procedure used for all targets.
    \emph{Top}:
    The source spectrum in a small wavelength range around \Ha\ (black),
    uncertainty in spectral flux (gray error bars).
    Over-plotted is the line model + background computed from the median
    posterior PDF sample (dark green line) and the 15th to 85th percentile
    region computed from all posterior PDF samples (shaded green region).
    \emph{Bottom}:
    Source spectrum residuals computed by subtracting the median model (dark
    green line in top panel) from the source spectrum.
    The over-plotted line (orange) shows the Gaussian process background model
    computed from the median posterior PDF sample.
    \label{fig:Halpha-mcmc-fit}}
\end{figure}

\figurename~\ref{fig:Halpha-mcmc-corner} shows a corner plot of all projections
of these posterior PDF samples for a randomly-chosen source (HD 63408, in this
case).
The marginal posterior distribution over $x_0$ can be converted to a
distribution over radial velocity; in this case, the estimated (Gaussian
standard deviation) precision is $\approx 2.5~\kms$.
\figurename~\ref{fig:Halpha-mcmc-fit} shows the segment of the source spectrum
around \Ha\ used in the line modeling (black line) with uncertainties (gray
bars).
Top panel shows the linear polynomial + absorption line model computed for the
median sample (dark green line); shaded region shows the 15th to 85th
percentiles computed from all posterior samples.
Bottom panel shows the residuals of the data from the median linear polynomial +
absorption line model (black line) along with the Gaussian process computed from
the median parameters (orange curve) used to model other correlated structure in
the vicinity of \Ha\ (which can be interpreted as correlated noise).
These figures are produced for all source spectrum fits and visually inspected
to ensure that the MCMC sampling appears to have converged for all cases.

For each source spectrum, with the resulting posterior samples, we compute the
median and median absolute deviation (MAD) of the marginal distribution over
line centroid, $x_0$.
We use the MAD to estimate the standard deviation ($\sigma \approx 1.5\times{\rm
MAD}$, assuming the samples are Gaussian distributed) of the samples and find
that the typical velocity \emph{precision} from measuring the centroid of \Ha\
is between $5$--$10~\kms$.
We report the measured absolute radial velocities along with compiled sky
position, astrometry, and $G$-band magnitude from the \tgas\ catalog and \tmass\
$J$-band magnitudes for all of our comoving star targets in
\tablename~\ref{tbl:data-stars} available with this \documentname\ (see table
caption for a detailed description of the contents).

\section{Data products} \label{sec:table}

See tables below for full descriptions of the columns in the associated data
files.

\begin{table}[ht]
    \centering
    \caption{Description of columns in data table containing absolute radial
    velocity measurements and summary information from the \tgas\ catalog.}
    \label{tbl:data-stars}
    \begin{tabular}{l|l|l}
        \toprule
        Column Name    & Unit & Description\\
        \midrule
        \texttt{Oh17\_group\_id}  &      & Group ID from the candidate catalog (\citealt{Oh:2017})\\
        \texttt{tgas\_source\_id} &      & Unique source identifier from \tgas\\
        \texttt{name}             &      & HD, \project{Hipparcos}, or \project{Tycho-2} identifier\\
        \texttt{ra}               & deg  & Right ascension from \tgas\\
        \texttt{dec}              & deg  & Declination from \tgas\\
        \texttt{parallax}         & mas  & Parallax from \tgas\\
        \texttt{distance}         & pc   & Distance estimate using inverse parallax from \tgas\\
        \texttt{G}                & mag  & \gaia\ $G$-band magnitudes\\
        \texttt{J}                & mag  & \tmass\ $J$-band magnitudes\\
        \texttt{rv}               & \kms & Radial velocity measured in this work\\
        \texttt{rv\_err}          & \kms & Radial velocity error\\
        \bottomrule
    \end{tabular}
\end{table}

\begin{table}[h]
    \centering
    \caption{Description of columns in data table containing separations and
    relative radial velocities for all observed groups, along with likelihood
    ratios computed with and without relative radial velocities.}
    \label{tbl:data-pairs}
    \begin{tabular}{l|l|l}
        \toprule
        Column Name    & Unit & Description\\
        \midrule
        \texttt{Oh17\_group\_id}    &      & Group ID from candidate catalog (\citealt{Oh:2017})\\
        \texttt{R\_mu}              &      & Log-likelihood-ratio computed with only astrometric information\\
        \texttt{R\_rv}              &      & Log-likelihood-ratio computed with radial velocities as well\\
        \texttt{sep\_tan}           & \pc  & Median of tangential separation error distribution\\
        \texttt{sep\_tan\_err}      & \pc  & Uncertainty in tangential separation\\
        \texttt{relative\_rv}       & \kms & Relative radial velocity between the two stars\\
        \texttt{relative\_rv\_err}  & \kms & Uncertainty in relative radial velocity\\
        \bottomrule
    \end{tabular}
\end{table}

\clearpage

\bibliographystyle{aasjournal}
\bibliography{refs}

\end{document}